\documentclass[
  superscriptaddress,
  amsmath,amssymb,
  onecolumn,
  notitlepage,
  aps,
  prd,
  longbibliography,
  nofootinbib,
  floatfix
]{revtex4-2}

\usepackage{graphicx,epsfig,amssymb} \usepackage{amsmath,amsfonts, times} 
\usepackage{bm} 
\usepackage{bbm} \usepackage{epstopdf} \usepackage[linktocpage,colorlinks]{hyperref} \usepackage[caption=false]{subfig} 
\usepackage[usenames]{color} \usepackage{natbib} \usepackage{soul} \usepackage[utf8x]{inputenc} \usepackage[usenames]{color} \usepackage{cancel} 
\usepackage{float}

\begin{document}

 \title{Circularly polarized gravitational waves from parity-violating scalar-tensor theory}

\author{Jia-Xi Feng}
\email{fengjx@ucas.ac.cn}
\affiliation{School of Fundamental Physics and Mathematical Sciences, Hangzhou Institute for Advanced Study, UCAS, Hangzhou 310024, China}
\affiliation{School of Physics and Astronomy, Sun Yat-sen University, Zhuhai 519082, China}
%
\author{Jia-Yuan Fang}
\email{fangjy37@mail2.sysu.edu.cn}
\affiliation{School of Physics and Astronomy, Sun Yat-sen University, Zhuhai 519082, China}
\author{Xian Gao}
\email{Corresponding author: gaoxian@mail.sysu.edu.cn}
\affiliation{School of Physics, Sun Yat-sen University, Guangzhou 510275, China}
\affiliation{Guangdong Provincial Key Laboratory of Quantum Metrology and Sensing, Sun Yat-sen University, Zhuhai 519082, China}

\begin{abstract}
We study both primordial gravitational waves (GWs) and scalar- induced gravitational waves (SIGWs) in a  class of the parity-violating scalar-tensor (PVST) theory, of which the Lagrangian is the linear combination of seven ghost-free parity-violating scalar-tensor monomials dubbed the ``Qi-Xiu'' Lagrangians. 
At linear order, we obtain the quadratic action for tensor perturbations and show that parity-violating terms associated with $\mathcal{L}_{1,2,5,6,7}$ render the tensor propagation polarization dependent, leading to chiral primordial spectra and a nonvanishing degree of circular polarization.
At second order, we derive the equation of motion for SIGWs and identify the explicit parity-violating source terms.
In particular, $\mathcal{L}_3$ and $\mathcal{L}_4$ enter exclusively through the source term for SIGWs, allowing parity violation to arise even when the linear GWs propagation remains effectively general-relativity-like.
During the radiation-dominated era,
we compute the fractional energy density of SIGWs for both monochromatic and log-normal curvature power spectra. We find that, around the peak frequency, SIGWs in PVST gravity exhibit characteristic deviations from those in general relativity, resulting in a nonzero degree of circular polarization.
\end{abstract}

\maketitle

\section{Introduction}
The detection of gravitational waves (GWs) by the Laser Interferometer Gravitational-Wave Observatory  scientific
collaboration and Virgo collaboration \cite{LIGOScientific:2016aoc,LIGOScientific:2017vwq,LIGOScientific:2018mvr,LIGOScientific:2020ibl,LIGOScientific:2021usb,KAGRA:2021vkt} not only opened a new window to probe the nature of gravity, but also ushered in the era of multimessenger astronomy. In the early Universe, gravitational waves could be generated by a variety of processes. In particular, primordial GWs produced during the inflationary epoch carry rich information about the early universe, although they have not been detected on cosmic microwave background scales. Additionally, scalar-induced gravitational waves (SIGWs), generated at second order through nonlinear scalar-tensor interactions, have attracted considerable attention in recent years \cite{Ananda:2006af,Baumann:2007zm,Saito:2008jc,Kohri:2018awv,Espinosa:2018eve,Domenech:2021ztg,Yuan:2021qgz,Inomata:2025wiv}. Notably, pulsar timing array (PTA) collaborations have reported  indications of a stochastic signal in the nanohertz band \cite{NANOGrav:2023gor,NANOGrav:2023hde,Zic:2023gta,EPTA:2023sfo,EPTA:2023fyk,Xu:2023wog}, and SIGWs have been discussed as a possible cosmological origin of the PTA signal \cite{NANOGrav:2023hvm,EPTA:2023xxk,Cai:2023dls,Yi:2023mbm,Yi:2023tdk,Wang:2023ost,Liu:2023ymk,Liu:2023hpw,Chen:2024fir}.
With planned GW observatories such as
LISA \cite{Danzmann:1997hm,LISA:2017pwj},  Taiji \cite{Hu:2017mde,Ruan:2018tsw}, TianQin \cite{TianQin:2015yph,Gong:2021gvw}, and DECIGO \cite{Kawamura:2020pcg}, as well as the Einstein
Telescope \cite{ET:2019dnz}, we are entering an era of GW astronomy.

In light of these advances, we use GWs as a probe of possible parity violation in the Universe. Parity violation is firmly established in the weak interaction \cite{Lee:1956qn,Wu:1957my}. Recently, cosmological observations have provided  possible indications of parity violation in the Universe, e.g., the galaxy trispectrum and the cross-correlation of $E$ mode and $B$ mode polarization \cite{Minami:2020odp,Eskilt:2022cff,Philcox:2022hkh,Hou:2022wfj}, which further motivate exploring whether parity violation can also arise in the gravitational interaction. From the observational perspective,  GW measurements offer an independent and complementary avenue, in particular, the cross-correlation between LISA and Taiji in a heliocentric network makes it possible to constrain the degree of circular polarization in the stochastic gravitational wave background (SGWB) \cite{Seto:2020zxw,Orlando:2020oko}. From a theoretical standpoint, incorporating parity-violating (PV) terms in the gravitational action opens new avenues for probing the fundamental structure of gravitational interactions. A variety of PV theories of gravity have been proposed and extensively studied, despite differing motivations. Within the Riemannian framework, the simplest realization is Chern-Simons (CS) gravity \cite{Jackiw:2003pm}; beyond CS gravity, further PV scenarios include Hořava gravity \cite{Horava:2009uw}, PV higher derivative gravity \cite{Crisostomi:2017ugk}, and PV spatially covariant gravity \cite{Gao:2019liu,Hu:2021bbo,Hu:2021yaq}. Extensions have also been considered in non-Riemannian frameworks, for example in PV teleparallel gravity \cite{Cai:2021uup,Wu:2021ndf,Li:2020xjt,Li:2021wij,Rao:2021azn,Li:2021mdp,Li:2022mti,Li:2022vtn,Hohmann:2022wrk,Conroy:2019ibo,Chen:2022wtz}.
GWs offer a particularly  promising avenue to test PV gravity. At linear order, GWs can exhibit characteristic PV signatures such as amplitude and velocity birefringence \cite{Satoh:2007gn,Saito:2007kt,Yunes:2010yf,Takahashi:2009wc,Myung:2009ug,Wang:2012fi,Zhu:2013fja,Cannone:2015rra,Zhao:2019szi,Zhao:2019xmm,Qiao:2019hkz,Qiao:2019wsh,Qiao:2021fwi,Gong:2021jgg,Qiao:2022mln,Qiao:2023hlr,Qiao:2025cmy,Zhang:2025kcw,Guo:2025bxz,Jia:2026vqo}. At second order, circularly polarized SIGWs were first studied in CS gravity \cite{Zhang:2022xmm,Feng:2023veu}, and subsequently extended to PV teleparallel gravities \cite{Zhang:2023scq,Zhang:2024vfw,Zhang:2025mps}, as well as to chiral scalar-tensor theories of gravity \cite{Feng:2024yic}.\footnote{SIGWs in other modified gravity theories have also been studied \cite{Domenech:2024drm, Zhou:2024doz,Jiang:2025ysb, Lopez:2025gfu,Domenech:2025zvi}.}

Recently, a systematic framework for parity-violating scalar-tensor (PVST) gravity has been constructed in \cite{Hu:2024hzo}.
Within this framework, the Lagrangians are polynomials built of the scalar field and its covariant derivatives coupled with the curvature tensor. A complete classification of the monomials was performed according to the total number of derivatives $d$ in each monomial in the unitary gauge with $\varphi=t$. Up to $d=4$, seven independent parity-violating scalar-tensor Lagrangians that are ghost-free were identified and dubbed the Qi-Xiu~\footnote{Qi-Xiu stands for ``seven constellations'' in classical Chinese \cite{Hu:2024hzo}.} Lagrangians.
In this work, we study the linear GWs and SIGWs within this PVST framework proposed in \cite{Hu:2024hzo}.
We first analyze linear tensor perturbations and derive the quadratic action for the tensor modes, demonstrating that the PV Lagrangians associated with $\mathcal{L}_{1,2,5,6,7}$ modify the propagation of linear tensor modes, and give rise to chiral primordial GWs during inflation, accompanied by a nonzero degree of circular polarization.
We then derive, for the first time, to the best of our knowledge, the equations of motion (EOM) for SIGWs in PVST gravity and obtain the explicit PV contributions to the source terms.
A particularly distinctive feature of the Qi-Xiu Lagrangians is that two of them, $\mathcal{L}_3$ and $\mathcal{L}_4$, do not modify the linear GWs but instead contribute exclusively to the source term of SIGWs.
As a result, PV effects can still be imprinted in SIGWs as a characteristic prediction of the theory even when the propagation of linear GWs is essentially indistinguishable from that in general relativity (GR).
Focusing on SIGWs during the radiation-dominated era, we compute the fractional energy density and the degree of circular polarization for both monochromatic and log-normal curvature power spectra, and investigate the characteristic signatures arising from PV source terms.

The paper is organized as follows.
In Sec.~\ref{Sec2}, we review the basic framework of the parity-violating scalar-tensor theory.
In Sec.~\ref{Sec3}, we derive the quadratic action for linear tensor perturbations and discuss the power spectrum and degree of circular polarization for primordial GWs.
In Sec.~\ref{Sec4}, we derive the EOM for SIGWs in the PVST theory and identify the PV source terms, highlighting the contributions from $\mathcal{L}_3$ and $\mathcal{L}_4$.
In Sec.~\ref{Sec5}, we analyze SIGWs during the radiation-dominated era and compute the  energy density and degree of circular polarization.
Finally, our conclusions are summarized in Sec.~\ref{Sec6}.
Additional derivations and information are provided in the Appendixes.

\section{The parity-violating
scalar-tensor gravity}\label{Sec2}

In this section, we briefly review the general framework of PVST gravity established in \cite{Hu:2024hzo}. The action for the PVST gravity model considered in this work is given by
 \begin{align}
 \label{action}
 S=\frac{1}{2\kappa^2}\int \mathrm{d}^4x\sqrt{-g}(R+\mathcal{L}_{\mathrm{PV}})+\int \mathrm{d}^4x\sqrt{-g}\mathcal{L}_{\varphi},   
 \end{align}
where $\kappa^2=8\pi G$, $\mathcal{L}_{\mathrm{PV}}$ denotes the PV Lagrangian, and $\mathcal{L}_{\varphi}$ denotes the Lagrangian for the scalar field, which typically takes the form
\begin{align}
\mathcal{L}_{\varphi} = -\frac{1}{2}g^{\mu\nu}\partial_\mu\varphi\partial_\nu\varphi-V(\varphi).
\end{align}
In the PVST gravity~\cite{Hu:2024hzo}, there exist seven independent and ghost-free PV Lagrangians, named Qi-Xiu,
\begin{equation}
\mathcal{L}_{\mathrm{PV}}=\sum^7_{n=1}b_nL_n
\end{equation}
where $b_n$ are coupling functions.
For $d=3$, \footnote{In \cite{Hu:2024hzo}, the Lagrangians are classified by an integer $d$, where $d$ stands for the total number of derivatives of the corresponding Lagrangian in the so-called unitary gauge. In the covariant formulation, $d$ can be understood as the energy dimension of the Lagrangians.} there is a single generally covariant scalar-tensor monomial of the $(1,1,0)$ category, i.e., linear in both the curvature tensor and the second derivative of the scalar field.
For $d=4$, there are five combinations of the $(1,2,0)$ category, i.e., linear in the curvature tensor and quadratic in the second derivatives of $\varphi$. There is also a combination of the $(2,0,0)$ category, which is quadratic in the curvature tensor and contains no higher derivatives of $\varphi$.
\begin{itemize}
    \item[(i)] $d=3$, the $(1,1,0)$ category\\
    \begin{equation}
    L_1=\frac{1}{\sigma^3}\varepsilon_{abcd}R_{ef}^{\ \ cd}\varphi^a\varphi^e\varphi^{bf};
    \end{equation}
    \item[(ii)] $d=4$, the $(1,2,0)$ category\\
    \begin{align}
& L_2 =\frac{1}{\sigma^4}\varepsilon_{a b c d} R_{e f}{}^{c d} \varphi^a \varphi^e \varphi_m^b \varphi^{f m}, \\
& L_3 =\frac{1}{\sigma^6}\varepsilon_{a b c d} R_{e f}{ }^{c d} \varphi^m \varphi^n \varphi^e \varphi^a \varphi_m^f \varphi_n^b, \\
& L_4 =\frac{1}{\sigma^4} \varepsilon_{a b c d} R^{a m e n} \varphi^b \varphi^f \varphi_e^c \varphi_f^d\left(g_{m n}+\frac{1}{\sigma^2}\varphi_n \varphi_m\right), \\
& L_5 
=\frac{1}{\sigma^4}\varepsilon_{a b c d} R_{e f}{}^{c m} \varphi^a \varphi^e \varphi^{b f} \varphi^{d n}\left(g_{m n}+\frac{1}{\sigma^2}\varphi_m \varphi_n\right), \\
&L_6 =\frac{1}{\sigma^3} \varepsilon_{a b c d} R_{e f}{}^{ c d} \varphi^a \varphi^e \varphi^{b f} \varphi^{m n}\left(g_{m n}+\frac{1}{\sigma^2} \varphi_m \varphi_n\right) .
\end{align}
    \item[(iii)]  $d=4$, the $(2,0,0)$ category\\
    \begin{align}
L_7=\frac{1}{\sigma^2}\varepsilon_{a b c d} R_{e f}{ }^{c d} R^{a m e n} \varphi^b \varphi^f\left(g_{m n}+\frac{1}{\sigma^2}  \varphi_m \varphi_n\right),
\end{align}
\end{itemize}
where $\sigma=\sqrt{-\varphi_a\varphi^a}$.

In our model, the PV terms do not affect the background evolution or the linear scalar perturbations \footnote{The corresponding EOM are presented in Appendix~\ref{app-bf}.}. They contribute, however, to both linear and second-order tensor perturbations, leading to modified propagation of primordial GWs and additional source terms for SIGWs, as will be discussed in the following sections.

\section{Linear Tensor Perturbations}\label{Sec3}

\subsection{First-order tensor modes}
The perturbed metric in the Friedmann–Lemaître–
Robertson–Walker (FLRW) spacetime is given by
\begin{align}
\mathrm{d}s^2=-a^2(\eta) \mathrm{d}\eta^2+a^2(\eta)(\delta_{ij}+\gamma_{ij})\mathrm{d}x^i\mathrm{d}x^j,
\end{align}
The quadratic action for first-order tensor perturbation $\gamma_{ij}$ contains parity-conserving (PC) and PV parts
\begin{equation}
    \label{S1hh} S^{(2)}_{\gamma\gamma}=S^{({\mathrm{PC}})}_{\gamma\gamma}+S^{({\mathrm{PV}})}_{\gamma\gamma}
\end{equation}
where the PC part is the same as that in GR
\begin{equation}
S^{({\mathrm{PC}})}_{\gamma\gamma} =\frac{1}{2\kappa^2} \int \mathrm{d}^4 x ~\frac{a^2}{ 4}\left[ \left( \gamma_{i j}^{\prime}\right)^2-
 \left(\partial_k  \gamma_{i j}\right)^2\right],
\end{equation}
where prime denotes a derivative with respect to the conformal time $\eta$.
The PV part of the action is  obtained as
\begin{align}
S^{({\mathrm{PV}})}_{\gamma\gamma} & =\frac{1}{2\kappa^2} \int \mathrm{d}^4 x ~\frac{a^2}{ 4}\left[ \frac{c_1(\eta)}{a M_{\mathrm{PV}}} \epsilon^{i j k} \gamma_{il}^{\prime} \partial_j  {\gamma'_{k }}^{l }
+\frac{c_2(\eta)}{aM_{\mathrm{PV}}} \epsilon^{i j k} \partial^2  \gamma_{i l} \partial_j  \gamma_{k}^{ \ l}\right],
\end{align}
where 
\begin{equation}\label{c1}
    \begin{split}
         \frac{c_1(\eta)}{M_{\mathrm{PV}}} =
    -\frac{2b_1 \varphi'}{|\varphi'|}
    + \frac{1}{a}\left(4b_2\mathcal{H}+b_5\mathcal{H}+6b_6\mathcal{H}+2b_7\mathcal{H}\right),
    \end{split}
\end{equation}
\begin{equation}\label{c2}
     \frac{c_2(\eta)}{M_{\mathrm{PV}}}=~\frac{b_7'}{a},~~~~~~~~~~~~~~~~~~~~~~~~~~~
     ~~~~~~~~~~~~~~~~~~~~~
\end{equation}
and
\begin{align}
   \frac{c_1(\eta)-c_2(\eta)}{M_{\mathrm{PV}}} =&-\frac{2b_1 \varphi'}{|\varphi'|}
       + \frac{1}{a}\left(4b_2\mathcal{H}+b_5\mathcal{H}+6b_6\mathcal{H}
       -b_7'+2b_7\mathcal{H}\right).
\end{align}
Here $M_{\mathrm{PV}}$ denotes the characteristic energy scale of parity violation, and $c_1$ and $c_2$ are coefficients normalized by $M_{\mathrm{PV}}$.

Varying the action \eqref{S1hh} with respect to the tensor perturbation $\gamma_{ij}$, we obtain the corresponding EOM
\begin{equation}
\gamma_{i j}^{\prime \prime}+2 \mathcal{H} \gamma_{i j}^{\prime}-\partial^2 \gamma_{i j}+\frac{\epsilon^{ilk}}{a M_{\mathrm{PV}}}\partial_l\Big[ c_1\gamma_{jk}^{\prime \prime}+\left(c_1\mathcal{H}+c_1'\right) \gamma_{jk}^{\prime}-c_2 \partial^2  \gamma_{jk}\Big]= 0,
\end{equation}
After performing a Fourier decomposition, we expand the tensor perturbation as
\begin{equation}
    \gamma_{i j}(\eta, \boldsymbol{x})=\sum_{A=R,~L}\int \frac{  d^3 k}{(2 \pi)^{3}} ~\gamma_{\boldsymbol{k}}^A(\eta)  e_{i j}^A(\boldsymbol{k}) e^{i \boldsymbol{k} \boldsymbol{x}},
\end{equation}
where $A=R,L$ label the right- and left-handed polarization states, respectively.
Using the identity $\epsilon_{ilk}k_l e_{jk}^A = \mathrm{i}\lambda^A e_{ij}^A$ with $\lambda^R=1$ and $\lambda^L=-1$,
the evolution equation for each polarization mode becomes
\begin{equation}
\label{T1-EOM}
\gamma^{A''}_{\boldsymbol{k}}+(2+\nu_A)\mathcal{H}\gamma^{A'}_{\boldsymbol{k}}+(c^A_\text{T})^2k^2\gamma^{A}_{\boldsymbol{k}}=0,
\end{equation}
 where $\nu_A$ characterizes the modification to the amplitude damping, and $c^A_\text{T}$ denotes the  propagation speed of GWs, given by
\begin{equation}
\begin{split}
    \nu_{A}=& \frac{\lambda^A k\left(c_1 -c_1^{\prime}\mathcal{H}^{-1}\right) /\left(a M_{\mathrm{PV}}\right)}{1-\lambda^Ak c_1 /\left(a M_{\mathrm{PV}}\right)} ,\\
(c^A_\text{T})^2
=&1+\mu_A=\frac{1-\lambda^Akc_2/ (a M_{\mathrm{PV}})}{1-\lambda^A k c_1 /(a M_{\mathrm{PV}})}.
\end{split}
\end{equation}
It is evident that $\nu_A$ is solely dependent on $c_1$, which, in turn, involves the coefficients $b_1, b_2, b_5, b_6,$ and $b_7$ (originating from $\mathcal{L}_{1,2,5,6,7}$).
In contrast, the deviation of the propagation speed, described by $\mu_A\equiv(c^A_\text{T})^2-1  \propto (c_1 - c_2)$, also exhibits a dependence on the same set of coefficients.

It is important to note that the Lagrangians $\mathcal{L}_3$ and $\mathcal{L}_4$ do not affect the EOM for first-order tensor perturbation. However, they may influence second-order tensor modes, as will be discussed in Sec. \ref{Sec4}.

Note it is possible to have a combination of coefficients such that the linear GWs propagate in the speed of light. According to (21),
 requiring
 \begin{equation}
     \left(c_{\mathrm{T}}^{A}\right)^{2}=\frac{1-\lambda^{A}kc_{2}/(aM_{\mathrm{PV}})}{1-\lambda^{A}kc_{1}/(aM_{\mathrm{PV}})}=1,
 \end{equation}
implies $c_1=c_2$, that is,
\begin{equation}
   -\frac{2b_{1}\varphi'}{\left|\varphi'\right|}+\frac{1}{a}\left(4b_{2}\mathcal{H}+b_{5}\mathcal{H}+6b_{6}\mathcal{H}+2b_{7}\mathcal{H}\right)=\frac{b_{7}'}{a}. 
\end{equation}
If we require $c_{\mathrm{T}}=1$ for any $a(t)$,
 we must have
 \begin{equation}
-\frac{2b_{1}\varphi'}{\left|\varphi'\right|}=\frac{b_{7}'}{a},
 \end{equation}
and
\begin{equation}
4b_{2}\mathcal{H}+b_{5}\mathcal{H}+6b_{6}\mathcal{H}+2b_{7}\mathcal{H}=0,
\end{equation}
which are two constraint equations among five coefficients,
 leaving three free coefficients.
 Together with $b_{3}$ and $b_{4}$,
 we thus have a class of PV scalar-tensor Lagrangians controlled by five free coefficients,
 in which linear GWs propagate with $c_{\mathrm{T}}=1$. This is also consistent with the analysis in \cite{Gao:2019liu} (see eqs.
 (121)-(124)),
 in which two constraints among the coefficients of six independent PV spatially covariant gravity (SCG) monomials are found,
 yielding four SCG Lagrangians with $c_{\mathrm{T}}=1$.

\subsection{Polarized primordial gravitational waves}

In the study of primordial GWs, we consider tensor perturbations around a spatially flat FLRW background as introduced above. From the action \eqref{action} of our model, one finds that the PV terms do not contribute to the background equation and hence leave the background dynamics unchanged. During slow-roll inflation, we assume that the Universe is dominated by a scalar field $\varphi$ serving as the inflaton, so that the Friedmann equation and the evolution of the scalar field take the same form as in GR,
\begin{equation}
H^2 = \frac{8\pi G}{3} \rho, \quad \rho = \frac{1}{2} \dot\varphi^2 + V(\varphi), \quad \ddot\varphi + 3H\dot\varphi+ \frac{dV}{d\varphi} = 0,
\end{equation}
where dot denotes a derivative with respect to the cosmic time $t$.
Under slow-roll conditions $|\ddot\varphi| \ll |3H\dot\varphi|$, $\dot\varphi^2 \ll V(\varphi)$, one can define Hubble slow-roll parameters. 
\begin{equation}
    \epsilon_1=-\frac{\dot{H}}{H^2}, \quad \epsilon_2=\frac{d \ln \epsilon_1}{d \ln a}, \quad \epsilon_3=\frac{d \ln \epsilon_2}{d \ln a}.
\end{equation}

\subsubsection{The analytical solution of EOM for primordial GWs}

Primordial GWs correspond to tensor perturbations on a homogeneous and isotropic background, with their EOM given by Eq.~\eqref{T1-EOM}. To compute the power spectra, we adopt the uniform asymptotic approximation method \cite{Zhu:2013upa,Zhu:2013fha,Zhu:2016srz}  to obtain an analytical solution. For convenience, Eq.~\eqref{T1-EOM} can be rewritten as
\begin{equation}\label{T1-EOMv1}
\tilde{\gamma}^{A''}_{\boldsymbol{k}}+\left[\left(1+\mu_A\right) k^2-\frac{B^{A''}}{B^{A}}\right ]\tilde{\gamma}^{A}_{\boldsymbol{k}}=0,
\end{equation}
where $\tilde{\gamma}^{A}_{\boldsymbol{k}}=B^A\gamma^{A}_{\boldsymbol{k}}$ and
\begin{equation}
    B^A=a\sqrt{1-\frac{\lambda^Akc_1}{aM_{\mathrm{PV}}}}.
\end{equation}
For primordial GWs generated during inflation, we assume that the background satisfies the slow-roll conditions and that parity-violating effects represent small deviations from GR.  Under these assumptions,  expanding the  $B^{A''}/B^A$ and $\mu_A$ in terms of the slow-roll parameters, \eqref{T1-EOMv1} can be recast as
\begin{equation}
    \label{T1-EOMv2}
\tilde{\gamma}^{A''}_{\boldsymbol{k}}+\left[\left(1-\lambda^Ak\eta(c_1-c_2)\epsilon_{*}\right) k^2-\frac{1}{\eta^2}\left(v_t^2-\frac{1}{4}\right)-\frac{\lambda^A k c_1}{\eta}\epsilon_{*}\right ]\tilde{\gamma}^{A}_{\boldsymbol{k}}=0,
\end{equation}
with
\begin{equation}
v_t=\frac{3}{2}+3\epsilon_1+4\epsilon_1^2+4\epsilon_1\epsilon_2+\mathcal{O}(\epsilon^3),
\qquad 
\epsilon_{*}=\frac{H}{M_{\mathrm{PV}}},
\end{equation}
where $\epsilon_*=H/M_{\mathrm{PV}}$ measures the magnitude of parity-violating corrections during inflation.

Applying the uniform asymptotic approximation method, the approximate solution can be expressed in terms of Airy-type functions as \cite{Zhu:2013upa,Zhu:2016srz}
\begin{equation}
\tilde{\gamma}^A=\alpha_0\left(\frac{\xi}{g(y)}\right)^{1/4}\mathrm{Ai}(\xi)
+\beta_0\left(\frac{\xi}{g(y)}\right)^{1/4}\mathrm{Bi}(\xi),
\end{equation}
where $y\equiv -k\eta$, 
$\alpha_0$ and $\beta_0$ are integration constants, 
$\mathrm{Ai}(\xi)$ and $\mathrm{Bi}(\xi)$ are the Airy functions, and $\xi$ is a function of $y$ given by \cite{Zhu:2013upa,Zhu:2016srz}
\begin{align} \xi(y) &= \begin{cases} \left(-\frac{3}{2}\int^y_{y_0}\sqrt{g(\bar y)}d\bar y\right)^{2/3}, & y\leq y_0,\\[4pt] -\left(\frac{3}{2}\int^y_{y_0}\sqrt{-g(\bar y)}d\bar y\right)^{2/3}, & y\geq y_0.\\ \end{cases} \end{align}
Here, the function $g(y)$ is given by
\begin{equation} 
g(y)=\frac{v_t^2}{y^2}-1-\lambda^Ay(c_1-c_2)\epsilon_*+\frac{\lambda^Ac_1\epsilon_*}{y},
\end{equation}
and $y_0$ is the single turning point of $g(y)$ \cite{Qiao:2019hkz}, derived by solving the equation $g(y)=0$,
\begin{align}
y_0= -\frac{1}{3 \lambda^A\left(c_1-c_2\right) \epsilon_*}\left[1-2^{\frac13}\left(1+3 \left(c_1-c_2\right) c_1 \epsilon_*^2\right) / Y -2^{-\frac13} Y\right],
\end{align}
with
\begin{align}
Y & =\left(Y_1+\sqrt{-4\left(1+3 \left(c_1-c_2\right) c_1 \epsilon_*^2\right)^3+Y_1^2}\right)^{\frac13}~, \\
Y_1 & =-2+27 v_t^2\left(\lambda^A c_1-\lambda^A c_2\right)^2 \epsilon_*^2-9 \left(c_1-c_2\right) c_1 \epsilon_*^2~.
\end{align}
Imposing the adiabatic vacuum initial condition on the approximate analytic solution in the subhorizon limit $y\rightarrow\infty$ fixes the integration constants, yielding
\begin{equation}
\alpha_0=\sqrt{\frac{\pi}{2k}} \mathrm{e}^{\mathrm{i}\pi/4},\qquad
\beta_0=\mathrm{i}\sqrt{\frac{\pi}{2k}} \mathrm{e}^{\mathrm{i}\pi/4}.
\end{equation}

\subsubsection{The power spectra and degree of circular polarization   of primordial GWs}

With the approximate solution of primordial GWs derived above, the corresponding power spectra $\mathcal{P}_{\text{T}}^{R,L}$  in the superhorizon limit $y \to 0$ are given by
\begin{equation}
\mathcal{P}_{\text{T}}^{R}=\frac{2k^3}{\pi^2}\left|\frac{\tilde{\gamma}^{R}}{B^{R}}\right|^2,
\qquad
\mathcal{P}_{\text{T}}^{L}=\frac{2k^3}{\pi^2}\left|\frac{\tilde{\gamma}^{L}}{B^{L}}\right|^2 .
\end{equation}
In this limit, the approximate form of $\tilde{\gamma}^{A}$ reads
\begin{equation}
\tilde{\gamma}^{A}(y)\simeq\frac{\mathrm{i}}{\sqrt{2k}}
\left(\frac{1}{g(y)}\right)^{\frac14}
\exp\left(\int_{y}^{y_0}\sqrt{g(\bar{y})} \mathrm{d}\bar{y}\right).
\end{equation}
The resulting power spectrum for each polarization mode is then obtained as~\cite{Qiao:2019hkz}
\begin{equation}
\begin{split}
    \mathcal{P}_{\text{T}}^{A}=\frac{k^2}{\pi^2}\frac{y}{(B^A)^2v_t}\exp{\left(2\int_{y}^{y_0}\sqrt{g(\bar{y})} \mathrm{d}\bar{y}\right)}
    \simeq\frac{18H^2}{\pi^2\mathrm{e}^3}\exp{\left(\frac{\pi\lambda^A\epsilon_{*}(9c_2-c_1)}{16}\right)}
\end{split}
\end{equation}
which can be related to the standard GR result by
\begin{equation}
\mathcal{P}_{\text{T}}^{A}
=\frac{\mathcal{P}_{\text{T}}^{\mathrm{GR}}}{2}
\left[1+\frac{\pi\lambda^A}{16}(9c_2-c_1)\epsilon_*
+\frac12\times\frac{\pi^2}{16^2}(9c_2-c_1)^2\epsilon_{*}^2
+\mathcal{O}(\epsilon_{*}^3)\right],
\end{equation}
where
\begin{equation}
\mathcal{P}_{\text{T}}^{\mathrm{GR}}
=\frac{2k^3}{\pi^2}
\left(\left|\frac{\tilde{\gamma}^{R}_{\boldsymbol{k}}}{B^{R}}\right|^2
+\left|\frac{\tilde{\gamma}^{L}_{\boldsymbol{k}}}{B^{L}}\right|^2\right),
\end{equation}
and the combination of coupling coefficients is given by
\begin{align}
\frac{9c_2-c_1}{M_{\mathrm{PV}}} =\frac{2b_1 \varphi'}{|\varphi'|} - \frac{1}{a}\left(4b_2\mathcal{H}+b_5\mathcal{H}+6b_6\mathcal{H} -9b_7'+2b_7\mathcal{H}\right).
\end{align}

The degree of circular polarization \cite{Saito:2007kt,Gluscevic:2010vv} of primordial GWs, is defined as 
\begin{equation}
\begin{split}\label{eq:Pi_def}
\Pi&\equiv
\frac{\mathcal{P}_{\text{T}}^{R}-\mathcal{P}_{\text{T}}^{L}}
{\mathcal{P}_{\text{T}}^{R}+\mathcal{P}_{\text{T}}^{L}}
\simeq
\frac{\pi}{16}(9c_2-c_1)\epsilon_{*}
+\mathcal{O}(\epsilon_{*}^3)\\
&\simeq
\frac{\pi}{16}M_{\mathrm{PV}}
\left[\frac{2b_1\varphi'}{|\varphi'|}
-\frac{1}{a}\left(4b_2\mathcal{H}
+b_5\mathcal{H}
+6b_6\mathcal{H}
-9b_7'
+2b_7\mathcal{H}\right)\right]\epsilon_{*}
+\mathcal{O}(\epsilon_{*}^3).
\end{split}
\end{equation}

It is worth emphasizing that both the power spectra and the degree of circular polarization of primordial GWs depend on the Lagrangians $\mathcal{L}_1$, $\mathcal{L}_2$, $\mathcal{L}_5$, $\mathcal{L}_6$, and $\mathcal{L}_7$, while they are independent of $\mathcal{L}_3$ and $\mathcal{L}_4$.
In contrast, for SIGWs, the Lagrangians $\mathcal{L}_3$ and $\mathcal{L}_4$ contribute solely via the source terms, as will be shown explicitly in Sec.~\ref{Sec4}.

\section{The scalar induced Gravitational waves}\label{Sec4}

In the Newtonian gauge, the perturbed FLRW metric can be written as
\begin{align}\label{metric-SIGWs}
\mathrm{d}s^2=-a^2(\eta)(1+2\phi)\mathrm{d}\eta^2+a^2(\eta)\left((1-2\psi)\delta_{ij}+\frac{1}{2}h_{ij}\right)\mathrm{d}x^i\mathrm{d}x^j,
\end{align}
where $\phi$ and $\psi$ denote the first-order scalar perturbations, and $h_{ij}$ is the second-order tensor perturbation (corresponding to SIGWs).
Throughout this work, we neglect vector perturbations and set the anisotropic stress to zero. Moreover, we do not include linear tensor perturbations as sources in the present analysis; their contribution will be left for future work.

The scalar field $\varphi$ can be decomposed into a homogeneous background and a perturbation as
\begin{equation}
    \varphi(\eta, \mathbf{x})=\bar\varphi(\eta)+\delta \varphi(\eta, \mathbf{x}),
\end{equation}
where $\bar{\varphi}(\eta)$ denotes the background value, which depends only on the conformal time $\eta$, and $\delta\varphi(\eta,\mathbf{x})$ is the corresponding perturbation. For convenience, we will henceforth drop the overbar and use $\varphi$ to denote the background field.

\subsection{The quadratic  and cubic action}

Using the perturbed metric in \eqref{metric-SIGWs}, we derive the action for SIGWs. We compute the action up to cubic order in perturbations, which can be written as
\begin{align}
\label{S2hh}
S_{\mathrm{GW}} = S^{(2)}_{hh} + S^{(3)}_{ssh},
\end{align}
where $S^{(2)}_{hh}$ denotes the quadratic action for tensor perturbations, and $S^{(3)}_{ssh}$ denotes the cubic interaction term involving two scalar and one tensor perturbations.

Both actions contain PC and PV contributions,
\begin{align}
S^{(2)}_{hh}=S^{({\mathrm{PC}})}_{hh}+S^{({\mathrm{PV}})}_{hh}, \qquad \quad
S^{(3)}_{ssh}=S^{({\mathrm{PC}})}_{ssh}+S^{({\mathrm{PV}})}_{ssh},
\end{align}
\subsubsection{Parity-conserving contributions}
The PC parts coincide with those in GR. The quadratic PC action takes the standard form
\begin{equation}
S^{(\text{PC})}_{hh}=\frac{1}{2\kappa^2} \int \mathrm{d}^4 x ~\frac{a^2}{16}\left[(h'_{ij})^2-(\partial_kh_{ij})^2\right],
\end{equation}
and the cubic PC term describing the scalar-tensor interaction is given by
\begin{equation}
    S^{(\text{PC})}_{ssh}=\frac{1}{2\kappa^2} \int \mathrm{d}^4 x ~\frac{a^2}{2}\left(2\partial_i\psi\partial_j\psi+\kappa^2\partial_i\delta\varphi\partial_j\delta\varphi\right)h_{ij},
\end{equation}
Therefore, the PC dynamics are identical to those in GR.

\subsubsection{Parity-violating contributions}
The PV part of the quadratic action can be written as  
\begin{equation}
S^{({\mathrm{PV}})}_{hh}  =\frac{1}{2\kappa^2} \int \mathrm{d}^4 x ~\frac{a^2}{16}\left[ \frac{c_1(\eta)}{aM_{\mathrm{PV}}} \epsilon^{i j k} h_{il}^{\prime} \partial_j  {h'_{k }}^{l }
+\frac{c_2(\eta)}{aM_{\mathrm{PV}}} \epsilon^{i j k} \partial^2  h_{i l} \partial_j  h_{k}^{ \ l}\right],
\end{equation}
where $c_1$ and $c_2$ (arising from $\mathcal{L}_{1,2,5,6,7}$) are defined in Eqs.~\eqref{c1} and \eqref{c2}. 
The PV cubic  action, which describes the interaction between two scalar perturbations and one tensor mode, can be decomposed into seven contributions,
\begin{equation}
    S^{({\mathrm{PV}})}_{ssh}  = \sum^7_{n=1}      S^{({\text{PV,n}})}_{ssh}.
\end{equation}
The explicit expressions for $S^{(\mathrm{PV},n)}_{ssh}$ are shown in Appendix~\ref{spvn}.

\subsection{The EOM of SIGWs}

Varying the action \eqref{S2hh} with respect to the tensor perturbation $h_{ij}$ yields the EOM for SIGWs
\begin{equation}
h_{i j}^{\prime \prime}+2 \mathcal{H} h_{i j}^{\prime}-\partial^2 h_{i j}+\frac{\epsilon^{ilk}}{a M_{\mathrm{PV}}}\partial_l\Big[ c_1\mathcal{H}h_{jk}^{\prime \prime}+\left(c_1\mathcal{H}+c_1'\right) h_{jk}^{\prime}-c_2 \partial^2  h_{jk}\Big]=4\mathcal{T}_{ij}^{lm}\mathcal{S}_{lm},
\end{equation}
where $\mathcal{T}_{ij}^{lm}$ is 
the projection operator. The source term can be decomposed as
\begin{equation}
\begin{split}
 \mathcal S_{ij}=\mathcal S^{(\mathrm{GR})}_{ij}+\mathcal S^{(\mathrm{PV})}_{ij},
 \end{split}
\end{equation}
where $ \mathcal S^{(\mathrm{GR})}_{ij} $ corresponds to the contribution from the PC term,
\begin{equation}
\mathcal S^{(\text{GR})}_{ij}=2\partial_i\psi\partial_j\psi+\kappa^2\partial_i\delta\varphi\partial_j\delta\varphi,
\end{equation}
which coincides with the expression in GR. The term $\mathcal S^{(\mathrm{PV})}_{ij}$ denotes the contribution arising from the PV term,
\begin{equation}
    \begin{split}
     \mathcal    S^{({\mathrm{PV}})}_{ij}=&
     \frac{\epsilon^{kl}_{\ \ i}}{a^2}\Big\{ \partial_{\eta}^2
     \Big(\mathcal A_{\varphi\varphi}\partial_l\delta\varphi\,\partial_j\partial_k\delta\varphi+\mathcal A_{\varphi\psi}\partial_l\delta\varphi\,\partial_j\partial_k\psi\Big) +\partial_{\eta}\Big(\mathcal B_{\varphi\varphi}\partial_l\delta\varphi\,\partial_j\partial_k\delta\varphi
\\[2pt]
&
+\mathcal B_{\varphi'\varphi}\partial_l\delta\varphi'\,\partial_j\partial_k\delta\varphi
+\mathcal B_{\varphi\psi}\partial_l\delta\varphi\,\partial_j\partial_k\psi
+\mathcal B_{\varphi'\psi}\partial_l\delta\varphi'\,\partial_j\partial_k\psi
+\mathcal B_{\varphi\psi'}\partial_l\delta\varphi\,\partial_j\partial_k\psi'
\\[2pt]
&
+\mathcal B_{\psi\psi}\partial_l\psi\,\partial_j\partial_k\psi
+\mathcal B_{\delta X}\partial_l\delta \varphi\,\partial_j\partial_k\delta X
+\mathcal B_{\psi X}\partial_l\delta X\,\partial_j\partial_k\psi
\\[2pt]
&
+\mathcal G^{(2)}_{\varphi\varphi} \partial_l\partial_m\delta\varphi\,\partial_j\partial_k\partial^m\delta\varphi
+\mathcal G^{(5)}_{\varphi\varphi} \partial_m\!\left(\partial_l\partial^m\delta\varphi\,\partial_j\partial_k\delta\varphi\right)+\mathcal G^{(6)}_{\varphi\varphi}\partial_l\nabla^2\delta\varphi\,\partial_j\partial_k\delta\varphi
\Big)\\[2pt]
&
+\mathcal K_{\varphi\varphi}\nabla^2(\partial_l\delta\varphi\,\partial_j\partial_k\delta\varphi)
+\mathcal K_{\varphi'\varphi}\nabla^2(\partial_l\delta\varphi'\,\partial_j\partial_k\delta\varphi)
+\mathcal K_{\varphi\psi}\nabla^2(\partial_l\delta\varphi\,\partial_j\partial_k\psi)
\Big\},
    \end{split}
\end{equation}
where
$\delta X=(\varphi'\delta\varphi'-\psi\,\varphi'^2)/a^2$, and we have used $\phi=\psi$.
The time-dependent coefficients $\mathcal A,\mathcal B,\mathcal G,\mathcal K$ are given by
{\small\begin{align}
        \mathcal A_{\varphi\varphi} =&-\frac{b_1a}{\varphi'|\varphi'|}+\frac{2b_2\mathcal{H}}{\varphi'^2}+\frac{b_5\mathcal{H}}{2\varphi'^2}+\frac{3b_6\mathcal{H}}{\varphi'^2}, 
        \qquad
        \mathcal A_{\varphi\psi}=-\frac{2b_7}{\varphi'} ,\nonumber\\[2pt]
            \mathcal B_{\varphi\varphi}=&\Big(-\frac{b_1\varphi''a}{\varphi'^2|\varphi'|}+\frac{b_{1\varphi}a}{|\varphi'|}\Big)+\Big(\frac{b_2\mathcal{H}^2}{\varphi'^2}+\frac{b_2\varphi''^2}{\varphi'^4}
   -\frac{2b_{2\varphi}\mathcal{H}}{\varphi'}\Big)  
   +\Big(-\frac{b_3\mathcal{H}^2}{\varphi'^2}+\frac{2b_3\mathcal{H}\varphi''}{\varphi'^3}-\frac{b_3{\varphi''}^2}{\varphi'^4}\Big)
\nonumber\\
   &
-\Big(\frac{b_4\mathcal{H}^2}{2\varphi'^2}
-\frac{b_4\mathcal{H}\varphi''}{2\varphi'^3}\Big)
+\Big(\frac{b_5\mathcal{H}\varphi''}{2\varphi'^3}-\frac{b_{5\varphi}\mathcal{H}}{2\varphi'}\Big)+\Big(\frac{3b_6\mathcal{H}\varphi''}{2\varphi'^3}-\frac{3b_{6\varphi}\mathcal{H}}{2\varphi'}\Big)+\Big(\frac{b_7\mathcal{H}'}{\varphi'^2}-\frac{b_{7\varphi}\mathcal{H}^2}{\varphi'^2}\Big),\nonumber\\[2pt]
    \mathcal B_{\varphi'\varphi}=&\frac{b_1a}{\varphi'|\varphi'|}-\frac{2b_2\varphi''}{\varphi'^3}-2\Big(\frac{b_3\mathcal{H}}{\varphi'^2}
-\frac{b_3\varphi''}{\varphi'^3}\Big)
-\frac{b_4\mathcal H}{2\varphi'^2}
-\frac{b_5\mathcal H}{2\varphi'^2}-\frac{3b_6\mathcal H}{\varphi'^2},
\qquad
\mathcal B_{\varphi'\varphi'}=\frac{b_2}{\varphi'^2}-\frac{b_3}{\varphi'^2},\nonumber\\[2pt]
\mathcal B_{\varphi\psi}=&-\frac{3b_1a}{|\varphi'|}+\Big(\frac{2b_2\varphi''}{\varphi'^2}+\frac{6b_2\mathcal H}{\varphi'}\Big)
+2\Big(\frac{b_3\mathcal H}{\varphi'}-\frac{b_3\varphi''}{\varphi'^2}\Big)
+\frac{b_4\varphi''}{2\varphi'^2}
+\frac{2b_5\mathcal H}{\varphi'}+\frac{12b_6\mathcal H}{\varphi'}
+\Big(\frac{4b_7\mathcal H}{\varphi'}+b_{7\varphi}\Big),\nonumber\\[2pt]
\mathcal B_{\varphi'\psi}=&-\frac{2b_2}{\varphi'}+\frac{2b_3}{\varphi'}-\frac{b_4}{2\varphi'}, \qquad
\mathcal B_{\varphi\psi'}=\frac{2b_2}{\varphi'}+\frac{2b_5}{\varphi'}+\frac{3b_6}{\varphi'}+\frac{2b_7}{\varphi'},\nonumber\\[2pt]
\mathcal B_{\varphi X}=&\frac{b_{1X}a}{|\varphi'|}-\frac{2b_{2X}\mathcal H}{\varphi'}
-\frac{b_{5X}\mathcal H}{2\varphi'}-\frac{3b_{6X}\mathcal H}{\varphi'},
\qquad 
\mathcal B_{\psi X}=b_{7X},
\qquad
\mathcal B_{\psi\psi}=b_2-b_3+\frac12b_4+b_7,\nonumber\\[2pt]
\mathcal G^{(2)}_{\varphi\varphi}=&-\frac{b_2}{\varphi'^2},
\qquad
\mathcal G^{(5)}_{\varphi\varphi}=\frac{b_5}{2\varphi'^2},
\qquad
\mathcal G^{(6)}_{\varphi\varphi}=\frac{b_6}{\varphi'^2},
\nonumber\\[2pt]
\mathcal K_{\varphi\varphi}=&-\frac{b_1a}{\varphi'|\varphi'|}-\frac{2b_2\mathcal H}{\varphi'^2}
-\Big(\frac{b_4\mathcal{H}}{2\varphi'^2}
-\frac{b_4\varphi''}{2\varphi'^3}\Big)
+\frac{b_5\mathcal H}{\varphi'^2}+\frac{3b_6\mathcal H}{\varphi'^2},
\nonumber\\[2pt]
\mathcal K_{\varphi\psi}=&\frac{b_4}{2\varphi'}+\frac{b_7}{\varphi'},
\qquad
\mathcal K_{\varphi'\varphi}=-\frac{b_4}{2\varphi'^2}.
\end{align}}
Here $b_{n\varphi}=\partial b_n/\partial\varphi$ and  $b_{nX}=\partial b_n/\partial X$ with $X\equiv-\tfrac12\nabla_a\varphi\,\nabla^a\varphi$.

In Fourier space, the EOM for SIGWs can be written as
\begin{equation}
\label{TEOM}
\tilde{h}^{A''}_{\boldsymbol{k}}(\eta)+\left[\left(1+\mu_A\right) k^2-\frac{B^{A''}}{B^{A}}\right ]\tilde{h}^{A}_{\boldsymbol{k}}(\eta)=\frac{4B^{A}}{z^{A}} \mathcal{S}_{\boldsymbol k}^{A}(\eta),
\end{equation}
where we have defined $\tilde{h}^{A}_{\boldsymbol{k}}\equiv B^{A}h^{A}_{\boldsymbol{k}}$ with $B^{A}\equiv a\sqrt{z^{A}}$, and
\begin{align}
z^{A}= 1-\frac{k\lambda^A c_1}{aM_{\mathrm{PV}}} ,
\qquad
\mu_A
=\frac{1}{z^A}\frac{\lambda^A k(c_1-c_2)}{ a M_{\mathrm{PV}}}.
\end{align}
Here,  $z^A$ and $\mu_A$ depend on the PV parameters $c_1,c_2$, 
 which receive contributions from the Lagrangians $\mathcal{L}_1$, $\mathcal{L}_2$, $\mathcal{L}_5$, $\mathcal{L}_6$, and $\mathcal{L}_7$, while $\mathcal{L}_3$, $\mathcal{L}_4$ do not contribute.

Using the method of Green’s function, we obtain the solution to Eq.~\eqref{TEOM} in the form
\begin{equation}
\label{solutionh}
h^{A}_{\boldsymbol k}\left(\eta\right)=\frac{4}{B^A(k,\eta)}\int^{\eta} d\bar{\eta}~G^{A}_{\boldsymbol k}\left(\eta,\bar{\eta}\right)
\frac{ B^A(k,\bar \eta)}{z^{A}(k,\bar \eta)}\mathcal{S}_{\boldsymbol k}^{A},
\end{equation}
where the Green’s function $G^{A}_{\boldsymbol k}$ satisfies
\begin{equation}
    {G^{A''}_{\boldsymbol{k}}}(\eta,\bar \eta)+\left[\left(1+\mu_A\right)k^2-\frac{{B^{A''}}}{B^{A}}\right ]G^{A}_{\boldsymbol{k}}(\eta,\bar \eta)= \delta(\eta-\bar \eta).
\end{equation}

We have shown that the PV contribution from Lagrangians $\mathcal{L}_1$,
$\mathcal{L}_2$, $\mathcal{L}_5$, $\mathcal{L}_6$, and $\mathcal{L}_7$,  modify the propagation of SIGWs through corrections to the effective dispersion relation and to the canonical normalization of the tensor modes (encoded in $z^A$ and $B^A$), and they can also contribute to the source term for SIGWs.
In contrast, $\mathcal{L}_3$ and $\mathcal{L}_4$ enter solely through the source term for SIGWs. This feature distinguishes our setup from other PV gravity models studied previously~\cite{Zhang:2022xmm,Feng:2023veu,Zhang:2023scq,Zhang:2024vfw,Feng:2024yic,Zhang:2025mps}. 

Accordingly, in the following we restrict attention to the contributions from $\mathcal{L}_3$ and $\mathcal{L}_4$ only. In this case, the Green's function reduces to its standard form,
\begin{equation}
\label{GF}
    {G^{A''}_{\boldsymbol{k}}}(\eta,\bar \eta)+\left(k^2-\frac{a''}{a}\right )G^{A}_{\boldsymbol{k}}(\eta,\bar \eta)= \delta(\eta-\bar \eta).
\end{equation}
In a sense, restricting attention to the contributions of $\mathcal{L}_3$ and $\mathcal{L}_4$ while neglecting the other terms ($\mathcal{L}_1,
\mathcal{L}_2, \mathcal{L}_5, \mathcal{L}_6, \mathcal{L}_7$) renders the propagation of linear GWs indistinguishable from that in GR. Nevertheless, terms can still imprint parity-violating effects on SIGWs.

\section{SIGWs during radiation-dominated era}\label{Sec5}
During the radiation-dominated era, the background equation of state is given by
\begin{align}
    \omega=\frac{\bar{p}}{\bar{\rho}}=\frac{1}{3}.
\end{align}
Substituting this relation into the background equations \eqref{EOM01} and \eqref{EOM02} yields
\begin{align}
    a = a_0\eta,\quad \mathcal{H}=\frac{1}{\eta},\quad  \varphi'=\pm\frac{2}{\kappa}\eta^{-1}.
\end{align}

For later convenience, we express the scalar perturbations in terms of the primordial curvature perturbation and the associated transfer function,
\begin{align}
    &\psi_{\boldsymbol k}=\frac{2}{3}\zeta_{\boldsymbol k}T_{\psi}(x),
\quad 
\delta\varphi _{\boldsymbol k}=\frac{2}{3}\zeta_{\boldsymbol k}\left(\frac{T'_{\psi}(x)+\mathcal{H}T_{\psi}(x)}{\mathcal{H}^2-\mathcal{H}'}\right)\varphi',
\end{align}
where $x=k\eta$, and  $\psi_{\boldsymbol k}$ and $\delta \varphi_{\boldsymbol k}$ denote the Fourier modes of the corresponding perturbations.

It is worth emphasizing that PV terms do not affect the background evolution nor the dynamics of scalar perturbations at linear order. Consequently, the transfer function $T_{\psi}(x)$ is unchanged from its GR form \cite{Kohri:2018awv}
\begin{equation}
\label{Tp}
T_{\psi}(x)= \frac{9}{x^2}\left(\frac{\sin(x/\sqrt{3})}{x/\sqrt{3}}-\cos(x/\sqrt{3})\right).
\end{equation}
As discussed at the end of the previous section, the Green's function governing tensor propagation reduces to its GR form,
\begin{align}
    G^A_{\boldsymbol k}(\eta, \bar{\eta})=\frac{\sin[k(\eta-\bar{\eta})]}{k}\Theta(\eta-\bar{\eta}).
\end{align}

\subsection{The power spectra}

The power spectra $\mathcal{P}^{A}_{h}(k,\eta)$ are related to the expectation values via
\begin{equation}
\label{hh}
   \left \langle h^A_{\boldsymbol{k}}(\eta) h^{A'}_{\boldsymbol{k}'}(\eta)\right\rangle =(2\pi)^3\delta^3(\boldsymbol k+\boldsymbol k')\frac{2\pi^2}{k^3}\delta^{AA'}\mathcal{P}^{A}_{h}(k,\eta).
\end{equation}
After some lengthy but straightforward calculations, the power spectra of the SIGWs can be expressed as \footnote{Hereafter, we consider only the PV contributions arising from $\mathcal{L}_3$ and $\mathcal{L}_4$, which enter the EOM solely through the source term for  SIGWs.}
\begin{align}
\mathcal{P}^{A}_h(k,\eta)=4\int_{0}^\infty\mathrm{d}v\int_{|1-v|}^{1+v}\mathrm{d}u
\left[\frac{4v^2-(1+v^2-u^2)^2}{4vu}\right]^2  I^{A}(k, u,v,x)^2\mathcal{P}_{\zeta}(u k)\mathcal{P}_{\zeta}(v k),
\end{align}
where $v=p/k$, $u=|{\boldsymbol k}-{\boldsymbol p}|/k$, and
\begin{equation}\label{IA}
    \begin{split}
I^{A}(k, u,v,x)
=&I_{\mathrm{GR}}(u,v,x)+I^{A}_{\mathrm{PV}}(k, u,v,x) \\
=&\int_{0}^x\mathrm{d}\bar{x}~k~\frac{a(\bar\eta)}{a(\eta)}G^A_{\boldsymbol k}(x,\bar{x})\left(
f_{\mathrm{GR}}(u,v,\bar x)
+f^A_{\mathrm{PV}}(k,u,v,\bar x) 
 \right),
    \end{split}
\end{equation}
where 
\begin{equation}\label{fGR}
    f_{\mathrm{GR}}(u,v,x)=\frac49\Big(2T_{\psi}(u x)T_{\psi}(v x)+ [T_{\psi}(u x)+u x T^{*}_{\psi}(u x)][T_{\psi}(v x)+v x T^{*}_{\psi}(v x)]\Big)
\end{equation}
Restricting to the PV effects sourced by $\mathcal{L}_3$ and $\mathcal{L}_4$, we keep only $f_{\mathrm{PV3}}$ and $f_{\mathrm{PV4}}$, and neglect the remaining PV terms. In this case,
\begin{equation}
f_{\mathrm{PV}}(u,v,x) = f_{\mathrm{PV3}}(u,v,x) + f_{\mathrm{PV4}}(u,v,x),
\end{equation}
with
\begin{align}\label{fPV}
    f_{\mathrm{PV3}}(u,v,x)=&-\frac{2\lambda^Ak}{9a^2}\Bigg\{
\partial_{\eta}\Big[-b_3\left(ux T^{*}_{\psi}(ux)+T_{\psi}(ux)\right)\left(vx T^{*}_{\psi}(vx)+T_{\psi}(vx)\right)\nonumber\\
&  
-b_3\left(u^2 x^2 T^{**}_{\psi}(ux)+2u x T^{*}_{\psi}(ux)\right)\left(vx T^{*}_{\psi}(vx)+T_{\psi}(vx)\right)-b_3 T_{\psi}(ux) T_{\psi}(vx) 
\nonumber\\
& 
+2b_3\left(ux T^{*}_{\psi}(ux)+T_{\psi}(ux)\right)T_{\psi}(vx)
+b_3\left(u^2 x^2 T^{**}_{\psi}(ux)+2u x T^{*}_{\psi}(ux)\right)T_{\psi}(vx) 
\nonumber\\
& -\frac{b_3}{4}\left(u^2 x^2 T^{**}_{\psi}(ux)+2u x T^{*}_{\psi}(ux)\right)\left(v^2 x^2 T^{**}_{\psi}(vx)+2v x T^{*}_{\psi}(vx)\right) \Big]
+(u\leftrightarrow v)\Bigg\},
\end{align}
and
\begin{align}
    f_{\mathrm{PV4}}(u,v,x)=&-\frac{2\lambda^Ak}{9a^2}\Bigg\{\partial_{\eta}\Big[
-\frac{b_4}{4}\left(ux T^{*}_{\psi}(ux)+T_{\psi}(ux)\right)\left(vx T^{*}_{\psi}(vx)+T_{\psi}(vx)\right)
\nonumber\\
& -\frac{b_4}{8}\left(u^2 x^2 T^{**}_{\psi}(ux)+2u x T^{*}_{\psi}(ux)\right)\left(vx T^{*}_{\psi}(vx)+T_{\psi}(vx)\right) +\frac{  b_4}{2}T_{\psi}(ux) T_{\psi}(vx)
\nonumber\\
&
-\frac{  b_4}{4}\left(ux T^{*}_{\psi}(ux)+T_{\psi}(ux)\right)T_{\psi}(vx)
-\frac{ b_4}{4}T_{\psi}(ux)\left(v^2 x^2 T^{**}_{\psi}(vx)+2v x T^{*}_{\psi}(vx)\right)\Big]\nonumber\\
&
+\Big[-\frac{b_4 k x}{8}\left(u^2 x^2 T^{**}_{\psi}(ux)+2u x T^{*}_{\psi}(ux)\right)\left(vx T^{*}_{\psi}(vx)+T_{\psi}(vx)\right)\nonumber\\
&-\frac{b_4 k x}{4}\left(ux T^{*}_{\psi}(ux)+T_{\psi}(ux)\right)\left(vx T^{*}_{\psi}(vx)+T_{\psi}(vx)\right)\nonumber\\
&
+\frac{b_4 k x}{4} \left(ux T^{*}_{\psi}(ux)+T_{\psi}(ux)\right)T_{\psi}(vx)\Big]\Bigg\}+(u\leftrightarrow v),
\end{align}
where 
${}^*$ denotes the derivative with respect to the argument.

Substituting Eq. \eqref{Tp} into Eqs. \eqref{fGR}--\eqref{fPV},
we obtain 
\begin{align}
         f_{\mathrm{GR}}(u,v,x)=&\frac{12}{u^3 v^3 x^6}\left(18 u v x^2 \cos \frac{u x}{\sqrt{3}} \cos \frac{v x}{\sqrt{3}}+\left(54-6\left(u^2+v^2\right) x^2+u^2 v^2 x^4\right) \sin \frac{u x}{\sqrt{3}} \sin \frac{v x}{\sqrt{3}}\right. \nonumber\\
&\left.+2 \sqrt{3} u x\left(v^2 x^2-9\right) \cos \frac{u x}{\sqrt{3}} \sin \frac{v x}{\sqrt{3}}+2 \sqrt{3} v x\left(u^2 x^2-9\right) \sin \frac{u x}{\sqrt{3}} \cos \frac{v x}{\sqrt{3}}\right) ,
\end{align}
which agrees with the GR result of \cite{Kohri:2018awv}.

For the parity-violating correction $f_{\mathrm{PV}}(u,v,x)$, we focus on the contributions arising from $\mathcal{L}_3$ and $\mathcal{L}_4$, assuming constant couplings $b_3$ and $b_4$. The corresponding expressions read
\begin{align}
     f_{\mathrm{PV3}}(u,v,x)=&-\frac{\mathcal C_3\lambda^Ak^4}{3u v x^5}
     \left( 6u vx^2\cos \frac{u x}{\sqrt{3}} \cos \frac{v x}{\sqrt{3}} +(18-3(u^2+ v^2)x^2)\sin \frac{u x}{\sqrt{3}} \sin \frac{v x}{\sqrt{3}}\right.\nonumber\\
     & \left. +\sqrt{3} u x \left(v^2 x^2-6\right) \cos \frac{u x}{\sqrt{3}} \sin \frac{v x}{\sqrt{3}}+ \sqrt{3} v x\left(u^2 x^2-6\right) \sin \frac{u x}{\sqrt{3}} \cos \frac{v x}{\sqrt{3}}  \right), 
\end{align}
and
\begin{align}
     f_{\mathrm{PV4}}(u,v,x)=&-\frac{\mathcal C_4\lambda^Ak^4}{4u^3 v^3 x^5}
     \left( -2 u v x^2 \left(u^2v^2 +3(u^2+v^2)\right)\cos \frac{u x}{\sqrt{3}} \cos \frac{v x}{\sqrt{3}}\right.\nonumber\\
     &\left.+\left(-18(u^2+ v^2)+6 u^2v^2(x^2-2)+u^2v^2x^2(u^2+v^2) \right)\sin \frac{u x}{\sqrt{3}} \sin \frac{v x}{\sqrt{3}}\right.\nonumber\\
     & \left. +\sqrt{3} u x \left(6(u^2+v^2) -u^2v^2 (x^2-3)\right) \cos \frac{u x}{\sqrt{3}} \sin \frac{v x}{\sqrt{3}}\right.\nonumber\\
     &\left.+ \sqrt{3} v x \left(6(u^2+v^2) -u^2v^2 (x^2-3)\right)\sin \frac{u x}{\sqrt{3}} \cos \frac{v x}{\sqrt{3}}  \right), 
\end{align}
where $\mathcal C_3=b_3/a_0^2$ and $\mathcal C_4=b_4/a_0^2$.


\subsection{The fractional energy density  and  degree of circular polarization of the SIGWs}

The fractional energy density of SIGWs is given by
\begin{equation} \label{OGW}
    \begin{split}
\Omega_{\mathrm{GW}}(k,\eta)
&=\frac{1}{48}\left(\frac{k}{\mathcal{H}}\right)^2
\sum_{A=R,L}\overline{\mathcal{P}^A_h(k,\eta)} \\
&=\frac{1}{12}\int_{0}^\infty\mathrm{d}v\int_{|1-v|}^{1+v}\mathrm{d}u
\left[\frac{4v^2-(1+v^2-u^2)^2}{4vu}\right]^2 
\sum_{A=R,L}
\overline{\tilde{I}^{A}(k,u,v,x)^2}
\mathcal{P}_{\zeta}(uk)\mathcal{P}_{\zeta}(vk),
    \end{split}
\end{equation}
where an overline denotes a time average, and $\overline{\tilde{I}^{A}(k,u,v,x)^2}=\overline{I^{A}(k,u,v,x)^2}x^2$. The explicit expression for the kernel $I^{A}$, together with a detailed derivation, is presented in Appendix~\ref{app-kernel}.

The degree of circular polarization  for SIGWs is
\begin{equation}
\Pi\equiv
\frac{\overline{\mathcal{P}^R_h}-\overline{\mathcal{P}^L_h}}
{\overline{\mathcal{P}^R_h}+\overline{\mathcal{P}^L_h}},
\end{equation}
where the overline again denotes a time average.

To illustrate the features of SIGWs in our model, we will discuss two forms of the primordial power spectrum for scalar perturbations: a monochromatic spectrum and a log-normal spectrum.

\subsubsection{Monochromatic spectrum }
We consider a monochromatic curvature perturbation \cite{Ananda:2006af, Inomata:2016rbd},
\begin{equation}
\label{ps1}
\mathcal{P}_{\zeta}(k)=\mathcal{A}_{\zeta}\delta(\log(k/k_s)).
\end{equation}
where $k_s$ is the peak scale, and $\mathcal{A}_\zeta$ is the amplitude.
After straightforward calculations, the fractional energy density of SIGWs is found to be
\begin{equation}\label{ph}
\Omega_{\mathrm{GW}}(k)=\frac{\mathcal{A}^2_\zeta}{12 \tilde{k}^{2}}\left(\frac{4-\tilde{k}^2}{4}\right)^2 \sum_{A=R,L} \overline{\tilde{I}^A(k,\tilde{k}^{-1},\tilde{k}^{-1},x\rightarrow \infty)^2}\Theta(2-\tilde{k}),
\end{equation}
where $\tilde{k}=k/k_s$.
The degree of circular polarization of SIGWs can be expressed as
\begin{equation}
\label{PI}
\Pi=\frac{\mathcal{N}(k,\tilde{k}^{-1},\tilde{k}^{-1})}{\mathcal{M}(k,\tilde{k}^{-1},\tilde{k}^{-1})}  \Theta(2-\tilde{k}),
\end{equation}
where 
\begin{align}
\label{nN}
\mathcal{N}(k,u,v)&=\overline{I^R(k,u,v,x\rightarrow \infty)^2}-\overline{I^L(k,u,v,x\rightarrow \infty)^2},\\
\label{mM}
\mathcal{M}(k,u,v)&=\overline{I^R(k,u,v,x\rightarrow \infty)^2}+\overline{I^L(k,u,v,x\rightarrow \infty)^2}.
\end{align}

\begin{figure}[htp]
\centering
\includegraphics[width=0.46\textwidth]{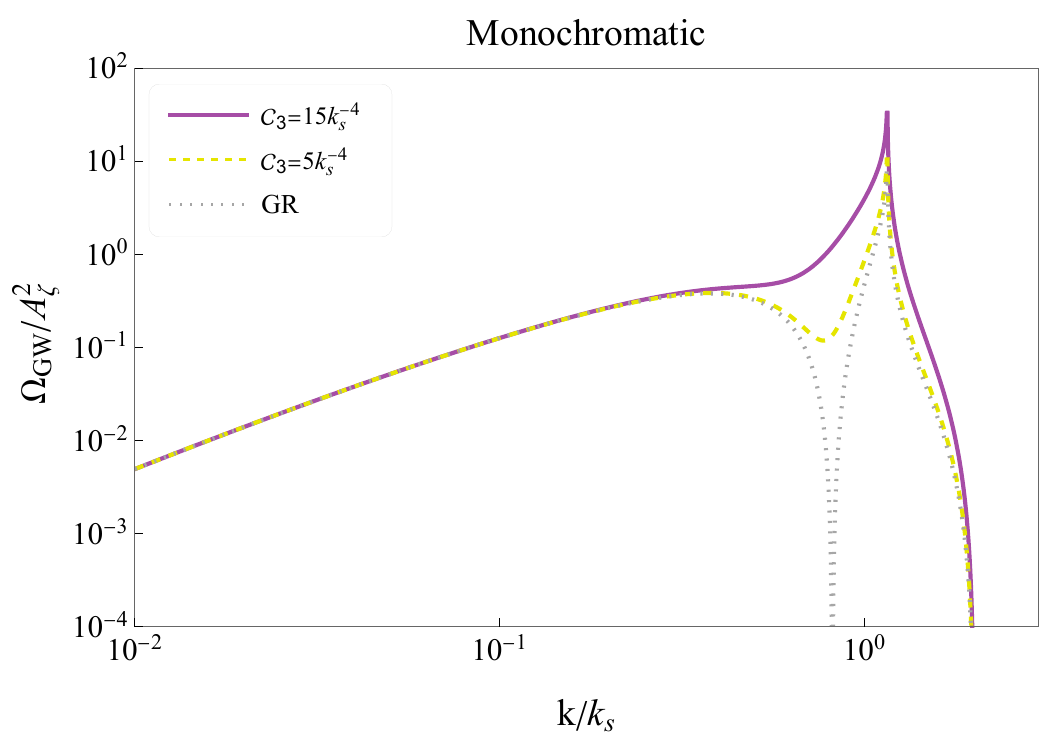}\hspace{0.2cm}
\includegraphics[width=0.46\textwidth]{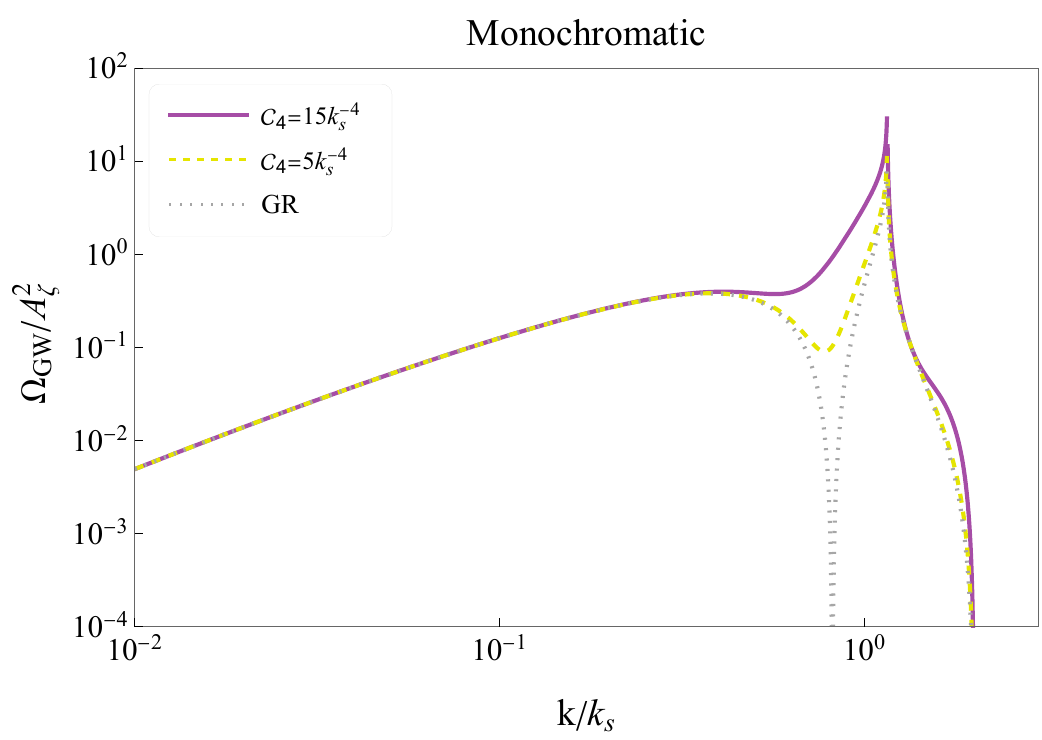}\hspace{0.2cm}
\caption{The fractional energy density $\Omega_{\mathrm{GW}}$ of SIGWs in GR and the PVST theory with a monochromatic curvature spectrum.}\label{Mfig1}
\end{figure}

\begin{figure}[htp]
\centering
\includegraphics[width=0.47\textwidth]{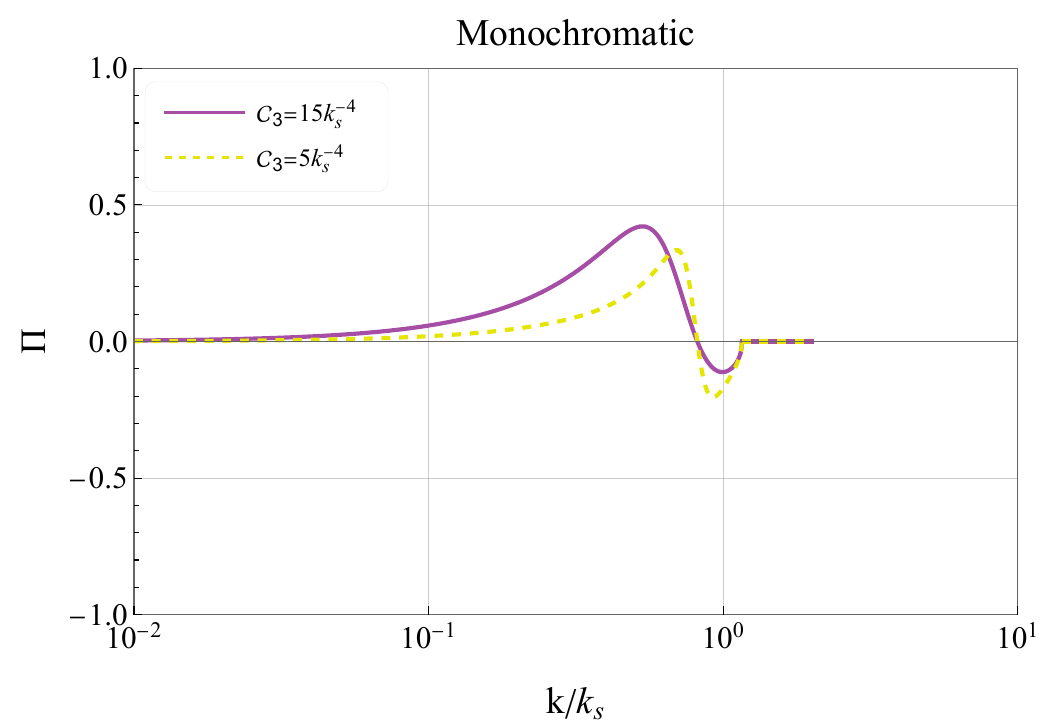}\hspace{0.3cm}
\includegraphics[width=0.47\textwidth]{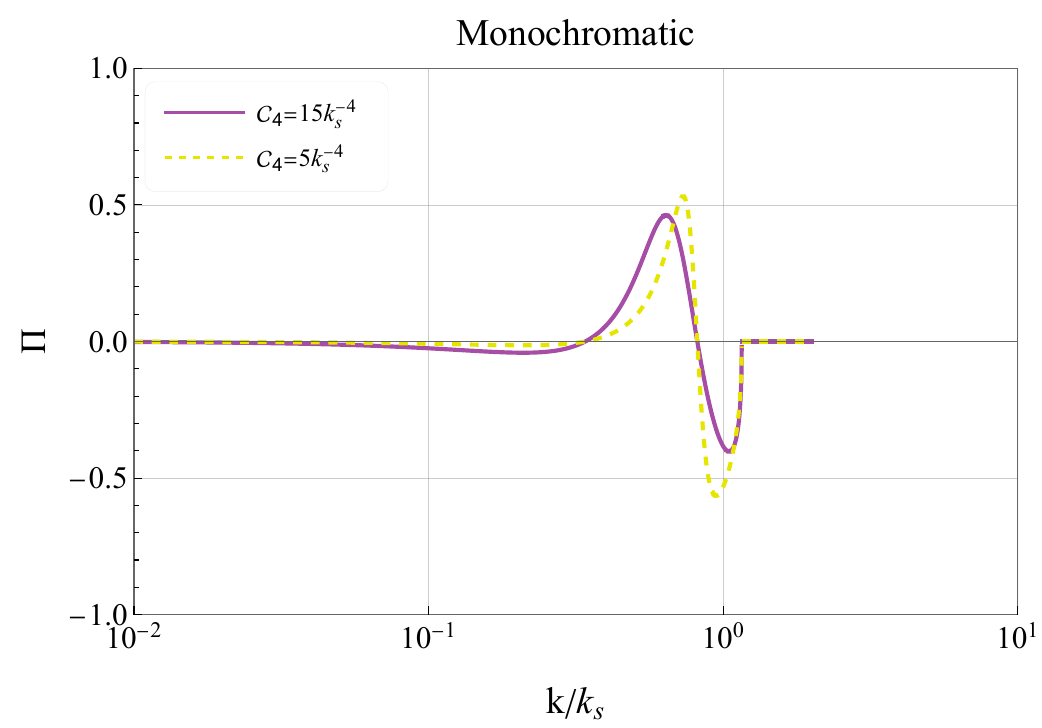}\hspace{0.3cm}
\caption{The degree of circular polarization of SIGWs with a monochromatic curvature spectrum. }\label{Mfig2}
\end{figure}

We compute the fractional energy density and the degree of circular polarization numerically for the spectrum \eqref{ps1}, and the results are shown in Figs.\ref{Mfig1} and \ref{Mfig2}.

Figure~\ref{Mfig1} shows the fractional energy density $\Omega_{\mathrm{GW}}$ of SIGWs in GR and the PVST theory, assuming a monochromatic curvature spectrum.
The left panel includes only the PV contribution from $\mathcal{L}_3$, while the right panel shows only the contribution from $\mathcal{L}_4$.

Compared with the GR prediction (gray dotted lines), the presence of the PV terms $\mathcal{L}_3$ or $\mathcal{L}_4$, slightly enhances the  $\Omega_{\mathrm{GW}}$ around the peak scale $k\simeq k_s$.
The magnitude of the deviation from GR depends on the strength of the coupling coefficients $\mathcal{C}_3$ and $\mathcal{C}_4$.
Larger coupling values produce a more pronounced enhancement near the peak, particularly around the resonance region $k=(2/\sqrt{3})k_s$.

Figure~\ref{Mfig2} shows the degree of circular polarization $\Pi$ of SIGWs in the PVST theory. The left panel includes only the PV contribution from $\mathcal{L}_3$, while the right panel shows only that from $\mathcal{L}_4$.

The PV terms $\mathcal{L}_3$ 
 and $\mathcal{L}_4$  lead to a nonzero degree of circular polarization, featuring a smooth peak with amplitude $\Pi \simeq 0.5$, thereby providing a clear signature of parity violation.

\subsubsection{The log-normal  spectrum}
To study a more realistic scenario, we consider a log-normal primordial spectrum for the curvature perturbation \cite{Espinosa:2018eve,Inomata:2016rbd},
\begin{equation}
\label{ps2}
\mathcal{P}_{\zeta}(k)=\frac{\mathcal{A}_\zeta}{\sqrt{2\pi}\sigma} \exp\left[-\frac{\log^2(k/k_s)}{2\sigma^2}\right],
\end{equation}
where  $\sigma$ characterizes the width of the peak. In the limit $\sigma\rightarrow 0$, the log-normal power spectrum reduces to the monochromatic form. For this spectrum, the energy density of  SIGWs remains finite as long as $\sigma$ is not extremely small, in contrast to the divergence that appears in the monochromatic limit.

\begin{figure}[htp]
\centering
\includegraphics[width=0.46\textwidth]{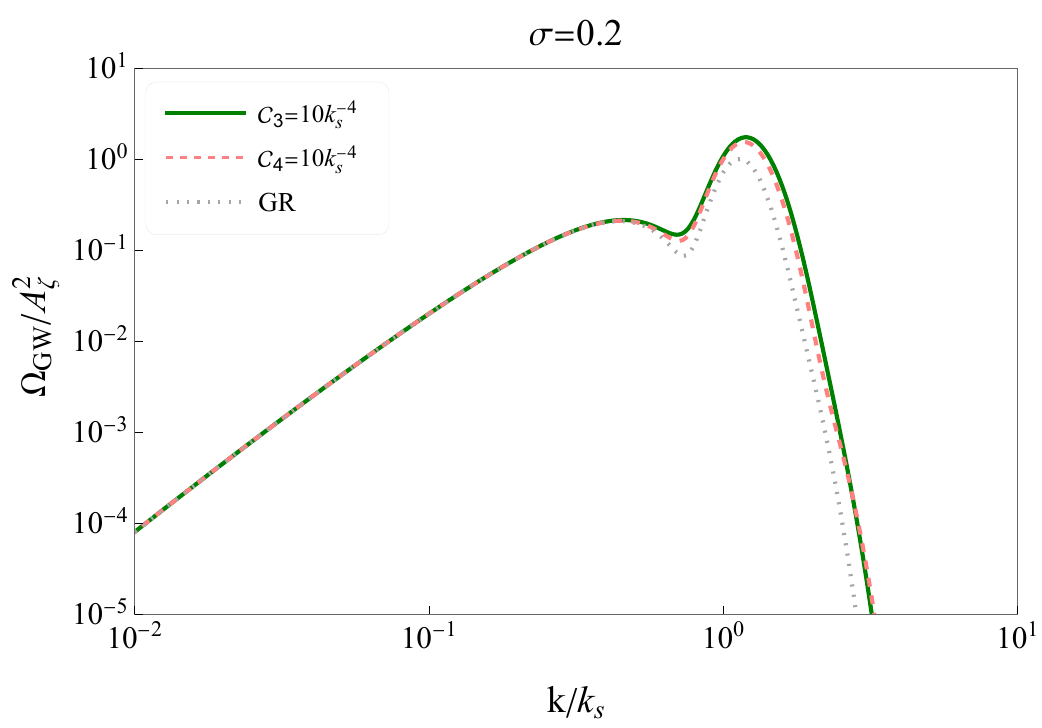}\hspace{0.2cm}
\includegraphics[width=0.46\textwidth]{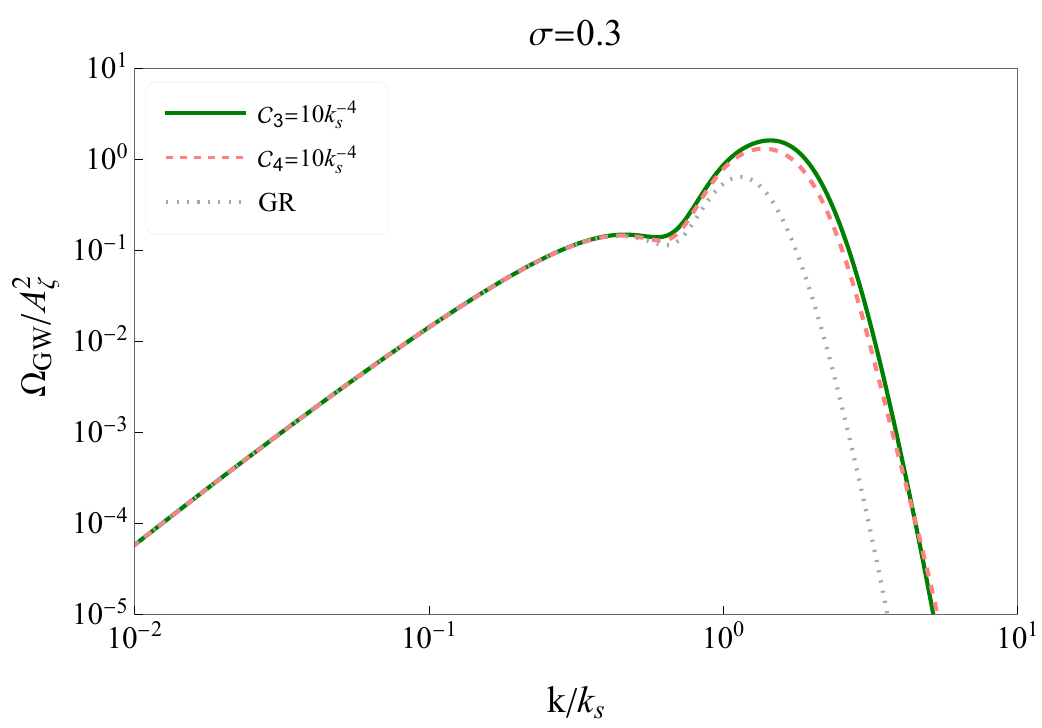}\hspace{0.2cm}
\caption{The fractional energy density $\Omega_{\mathrm{GW}}$ of SIGWs in GR and the PVST theory with a log-normal spectrum. The left panel corresponds to $\sigma=0.2$, while the right panel corresponds to $\sigma=0.3$.}\label{Lfig1}
\end{figure}
\begin{figure}[htp]
\centering
\includegraphics[width=0.46\textwidth]{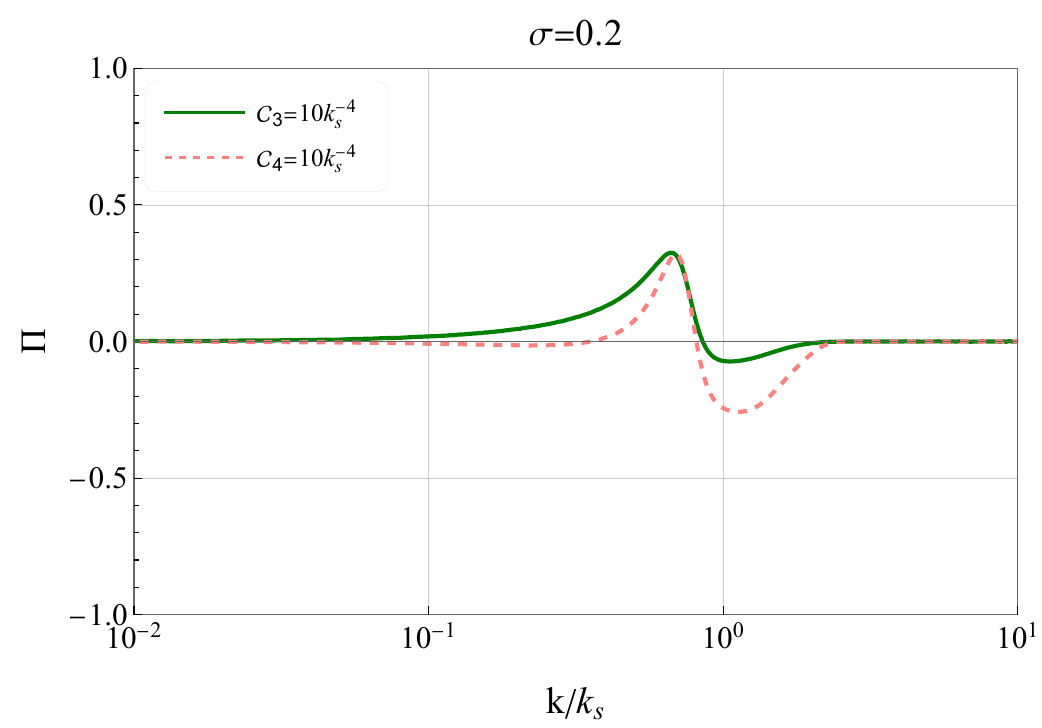}\hspace{0.2cm}
\includegraphics[width=0.46\textwidth]{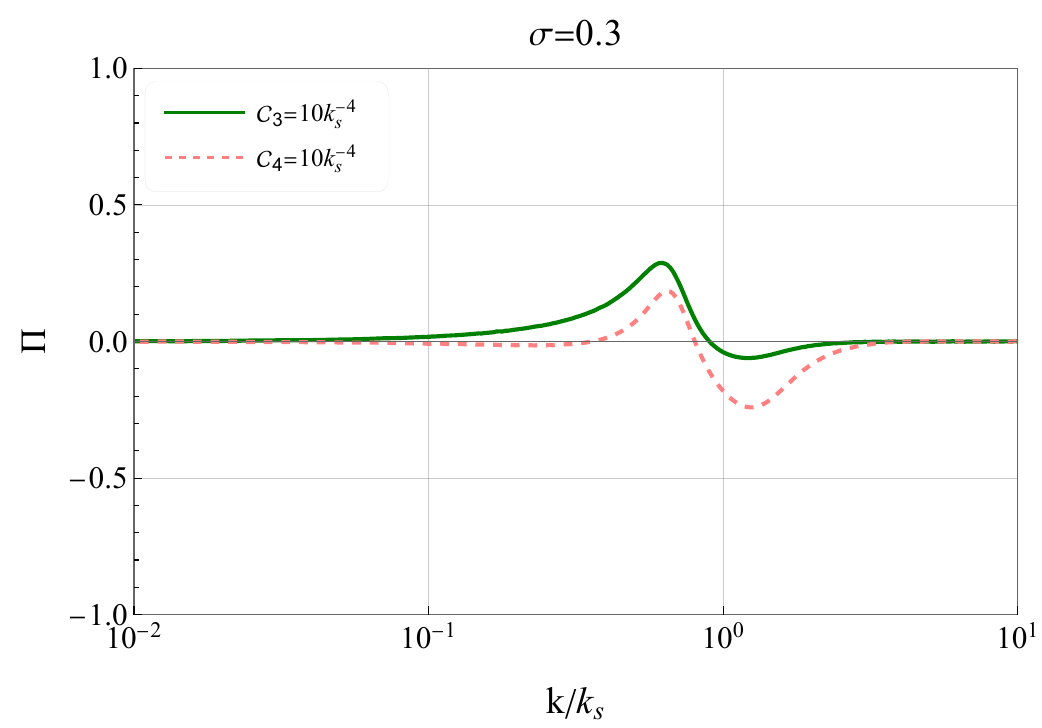}\hspace{0.2cm}
\caption{The degree of circular polarization of SIGWs with a log-normal spectrum. Left: $\sigma=0.2$.
Right: $\sigma=0.3$. }\label{Lfig2}
\end{figure}

We numerically compute the fractional energy density and the degree of circular polarization of SIGWs in our model with this log-normal peak \eqref{ps2}, and present the results for $\sigma=0.2$ and $\sigma=0.3$.
The numerical results are shown in  
Figs.~\ref{Lfig1}--\ref{LfinalPlotPI} (see Appendix~\ref{fig-log-normal} for Figs.~\ref{LfinalPlotOM} and \ref{LfinalPlotPI} ).


From Fig.~\ref{Lfig1}, we see that when the couplings associated with
$\mathcal{L}_3$ (green solid curves)
 and $\mathcal{L}_4$ (red dashed curves) are of the same order, the PV contribution from
$\mathcal{L}_3$
dominates near the peak scale, while $\mathcal{L}_4$ tends to exceed at scales beyond the peak. In addition, as $\sigma$ increases, the peak of the energy density becomes noticeably smoother.
The PV corrections from
$\mathcal{L}_3$ and $\mathcal{L}_4$
also induce a nonzero degree of circular polarization,  reflecting the PV nature of the underlying theory (see Figs. \ref{Lfig2} and \ref{LfinalPlotOM}). 
 
From Fig.~\ref{LfinalPlotPI}, we further observe that, as in the monochromatic case, increasing the coupling strengths $\mathcal{C}_3$ and $\mathcal{C}_4$ amplify the deviation of the resulting SIGW energy density from the GR prediction. Unlike the monochromatic spectrum, however, the peak remains finite provided that $\sigma$ is not exceedingly small, and becomes progressively smoother as $\sigma$ increases.

\section{Conclusion}\label{Sec6}

In this work, we  investigated  PV effects in GWs within the PVST framework of \cite{Hu:2024hzo}, in which seven independent parity-violating scalar-tensor terms that are free of ghosts (the Qi-Xiu Lagrangians) were identified. By analyzing both primordial GWs and SIGWs, we disentangle the contributions of the Qi-Xiu Lagrangians at linear order and second order in cosmological perturbation theory.

At linear order, we obtained the quadratic action for tensor perturbations and showed that the PV contributions encoded in $c_1$ and $c_2$ (originating from $\mathcal{L}_{1,2,5,6,7}$) modify the propagation of tensor modes.
As a result, primordial GWs become chiral, exhibiting unequal right- and left-handed power spectra and a nonzero degree of circular polarization, while the background evolution and the dynamics of linear scalar perturbations remain unchanged.

At second order, we derived the EOM for SIGWs in the PVST framework and obtained explicit expressions for the PV source terms.
A distinctive feature of the Qi-Xiu Lagrangians is the role of $\mathcal{L}_3$ and $\mathcal{L}_4$: they do not affect the quadratic tensor action, but affect the second-order tensor perturbation only through the source term that drives SIGWs.
Consequently, parity violation can be imprinted on SIGWs even when the  propagation of linear GWs retains its GR form.

To illustrate the phenomenology, we evaluated the fractional energy density of SIGWs and the degree of  circular polarization for
 a monochromatic spectrum and a log-normal spectrum.
In both cases, the contributions of $\mathcal{L}_3$ and $\mathcal{L}_4$  amplify  the deviation of the fractional energy density $\Omega_{\rm GW}$
from the GR prediction around the peak and generate a characteristic nonzero degree of circular polarization $\Pi$. For the monochromatic spectrum, the degree of circular polarization can reach $\Pi \simeq 0.5$ near the peak for typical parameter choices, and the effect grows with the couplings $\mathcal{C}_3$ and $\mathcal{C}_4$.
For the log-normal spectrum, the qualitative behavior remains similar, with the peak becoming smoother as the width parameter $\sigma$ increases.

These results highlight the rich phenomenology of PVST theory in the GW sector and clarify the hierarchy of contributions from the Qi-Xiu Lagrangians:  $\mathcal{L}_{1,2,5,6,7}$ modify tensor propagation already at linear order and also enter the second-order SIGWs dynamics, whereas $\mathcal{L}_3$ and $\mathcal{L}_4$ contribute only at second order as purely PV source terms for SIGWs. Looking ahead, since PVST theory generically predicts a nonvanishing degree of circular polarization, joint observations with future space-based detectors, in particular the cross-correlation between LISA and Taiji (or TianQin), will offer a promising avenue to search for (or constrain) parity violation in an isotropic SGWB.  A complete second-order treatment incorporating the full set of Qi-Xiu Lagrangians in the SIGW sector, including all source terms and mode couplings, is left for future work.

\begin{acknowledgments}
We would like to thank Prof. Tao Zhu and Fengge Zhang for helpful discussions and valuable suggestions. This work was partly supported by  National Natural Science Foundation of China (NSFC) under Grants No.  12547125, No. 12475068, and No. 11975020, and the Guangdong Basic and Applied Basic Research Foundation under Grant No. 2025A1515012977.
\end{acknowledgments}

\appendix
\section{$S^{({\mathrm{PV,n}})}_{ssh}$}\label{spvn}

The explicit expressions for $S^{({\mathrm{PV,n}})}_{ssh}$ are presented in the following 
\begin{align}
        S^{({\mathrm{PV1}})}_{ssh}=&\frac{1}{2\kappa^2} \int \mathrm{d}^4 x~\frac12\epsilon^{kl}_{ \ \ i}\left\{-\partial_{\eta}^2\left(\frac{b_1a}{\varphi'|\varphi'|}\partial_l\delta\varphi\partial_j\partial_k\delta\varphi\right) 
    +\partial_{\eta}\left[
\frac{b_1a}{\varphi'|\varphi'|}\partial_l\delta\varphi'\partial_j\partial_k\delta\varphi
\right.\right.
\nonumber\\
&\left.\left.
-\frac{b_1\varphi''a}{\varphi'^2|\varphi'|}\partial_l\delta\varphi\partial_j\partial_k\delta\varphi
-\frac{3b_1a}{|\varphi'|}\partial_l\delta\varphi\partial_j\partial_k\psi
+\frac{b_{1\varphi}a}{|\varphi'|}\partial_l\delta\varphi\partial_j\partial_k\delta\varphi
\right.\right.
\nonumber\\
&\left.\left.
+\frac{b_{1X}a}{|\varphi'|}\partial_l\delta X\partial_j\partial_k\delta\varphi \right]
-\frac{b_1a}{\varphi'|\varphi'|}\nabla^2(\partial_l\delta\varphi\partial_j\partial_k\delta\varphi)
\right\}h^{ij},
\end{align}
\begin{align}
        S^{({\mathrm{PV2}})}_{ssh}=&\frac{1}{2\kappa^2} \int \mathrm{d}^4 x~\frac12\epsilon^{kl}_{\ \ i}\left\{\partial_{\eta}^2\left(\frac{2b_2\mathcal{H}}{\varphi'^2}\partial_l\delta\varphi\partial_j\partial_k\delta\varphi\right)
        +\partial_{\eta}\left[\left(\frac{b_2\mathcal{H}^2}{\varphi'^2}+\frac{b_2\varphi''^2}{\varphi'^4}\right)\partial_l\delta\varphi\partial_j\partial_k\delta\varphi
       \right.\right.
       \nonumber\\
       &\left.\left.
       -\frac{2b_2\varphi''}{\varphi'^3}\partial_l\delta\varphi\partial_j\partial_k\delta\varphi' 
       +\frac{b_2}{\varphi'^2}\partial_l\delta\varphi'\partial_j\partial_k\delta\varphi' 
       +\frac{2b_2\varphi''}{\varphi'^2}\partial_l\delta\varphi\partial_j\partial_k\psi
        +\frac{6b_2\mathcal H}{\varphi'}\partial_l\delta\varphi\partial_j\partial_k\psi
        \right.\right.
        \nonumber\\
       &\left.\left.
       -\frac{2b_2}{\varphi'}\partial_l\psi\partial_j\partial_k\delta\varphi'
+\frac{2b_2}{\varphi'}\partial_l\psi'\partial_j\partial_k\delta\varphi
-\frac{2b_{2\varphi}\mathcal{H}}{\varphi'}\partial_l\delta\varphi\partial_j\partial_k\delta\varphi
 -\frac{2b_{2X}\mathcal{H}}{\varphi'}\partial_l\delta X\partial_j\partial_k\delta\varphi
 \right.\right.
\nonumber\\
&\left.\left.
  +b_2\partial_l\psi\partial_j\partial_k\psi    
  -\frac{b_2}{\varphi'^2}\partial_l\partial_m\delta\varphi\partial_j\partial_k\partial^m\delta\varphi\right]
+\frac{2b_2\mathcal{H}}{\varphi'^2}\nabla^2(\partial_l\delta\varphi\partial_j\partial_k\delta\varphi)\right\}h_{ij},
\end{align}  
\begin{align}
        S^{({\mathrm{PV3}})}_{ssh}=&\frac{1}{2\kappa^2} \int \mathrm{d}^4 x~\frac12\epsilon^{kl}_{\ \ i}\left\{
\partial_{\eta}\left[\left(-\frac{b_3\mathcal{H}^2}{\varphi'^2}+\frac{2b_3\mathcal{H}\varphi''}{\varphi'^3}-\frac{b_3{\varphi''}^2}{\varphi'^4}\right)\partial_l\delta\varphi\partial_j\partial_k\delta\varphi
 \right.\right.
        \nonumber\\
       &\left.\left.
-2\left(\frac{b_3\mathcal{H}}{\varphi'^2}
-\frac{b_3\varphi''}{\varphi'^3}\right)\partial_l\delta\varphi\partial_j\partial_k\delta\varphi'
+2\left(\frac{b_3\mathcal{H}}{\varphi'}
-\frac{b_3\varphi''}{\varphi'^2}\right)\partial_l\delta\varphi\partial_j\partial_k\psi
 \right.\right.
        \nonumber\\
       &\left.\left.
+\frac{2b_3}{\varphi'}\partial_l\psi\partial_j\partial_k\delta\varphi'
-\frac{b_3}{\varphi'^2}\partial_l\delta\varphi'\partial_j\partial_k\delta\varphi' 
-b_3\partial_l\psi\partial_j\partial_k\psi\right]\right\}h_{ij},
\end{align}
\begin{align}
            S^{({\mathrm{PV4}})}_{ssh}=&\frac{1}{2\kappa^2} \int \mathrm{d}^4 x~\frac12\epsilon^{kl}_{\ \ i}\left\{\partial_{\eta}\left[
\left(-\frac{b_4\mathcal{H}^2}{2\varphi'^2}
+\frac{b_4\mathcal{H}\varphi''}{2\varphi'^3}\right)\partial_l\delta\varphi\partial_j\partial_k\delta\varphi
-\frac{b_4\mathcal{H}}{2\varphi'^2}\partial_l\delta\varphi'\partial_j\partial_k\delta\varphi
 \right.\right.
        \nonumber\\
       &\left.\left.
+\frac{  b_4 \varphi''}{2\varphi'^2}\partial_l\delta\varphi\partial_j\partial_k\psi
+\frac{  b_4}{2}\partial_l\psi\partial_j\partial_k\psi
-\frac{ b_4}{2\varphi'}\partial_l\psi\partial_j\partial_k\delta\varphi'\right]
-\frac{b_4}{2\varphi'^2}\nabla^2(\partial_l\delta\varphi'\partial_j\partial_k\delta\varphi)
\right.
\nonumber\\
&\left.
-\left(\frac{b_4\mathcal{H}}{2\varphi'^2}
-\frac{b_4\varphi''}{2\varphi'^3}\right)\nabla^2(\partial_l\delta\varphi\partial_j\partial_k\delta\varphi)
+\frac{b_4}{2\varphi'}\nabla^2(\partial_l\psi\partial_j\partial_k\delta\varphi)\right\}h_{ij},
\end{align}
\begin{align}
        S^{({\mathrm{PV5}})}_{ssh}=&\frac{1}{2\kappa^2} \int \mathrm{d}^4 x~\frac12\epsilon^{kl}_{\ \ i}\left\{\partial_{\eta}^2\left(\frac{b_5\mathcal{H}}{2\varphi'^2}\partial_l\delta\varphi\partial_j\partial_k\delta\varphi\right)
+\partial_{\eta}\left[
\frac{b_5\mathcal{H}\varphi''}{2\varphi'^3}\partial_l\delta\varphi\partial_j\partial_k\delta\varphi
 \right.\right.
        \nonumber\\
       &\left.\left.
-\frac{b_{5\varphi}\mathcal{H}}{2\varphi'}\partial_l\delta\varphi\partial_j\partial_k\delta\varphi
-\frac{b_5\mathcal{H}}{2\varphi'^2}\partial_l\delta\varphi'\partial_j\partial_k\delta\varphi
+\frac{  2b_5 \mathcal{H}}{\varphi'}\partial_l\psi\partial_j\partial_k\delta\varphi
+\frac{ b_5}{2\varphi'}\partial_l\psi'\partial_j\partial_k\delta\varphi
 \right.\right.
        \nonumber\\
       &\left.\left.
-\frac{b_{5X}\mathcal{H}}{2\varphi'}\partial_l\delta X\partial_j\partial_k\delta\varphi+\frac{b_5}{2\varphi'^2}\partial_m\left(\partial_l\partial^m\delta\varphi\partial_j\partial_k\delta\varphi\right)\right]
+\frac{b_5\mathcal{H}}{\varphi'^2}\nabla^2(\partial_l\delta\varphi\partial_j\partial_k\delta\varphi)
\right\}h_{ij},
\end{align}
\begin{align}
        S^{({\mathrm{PV6}})}_{ssh}=& \frac{1}{2\kappa^2} \int \mathrm{d}^4 x~\frac12\epsilon^{kl}_{\ \ i}\left\{\partial_{\eta}^2\left(\frac{3b_6\mathcal{H}}{\varphi'^2}\partial_l\delta\varphi\partial_j\partial_k\delta\varphi\right)
+\partial_{\eta}\left[
\frac{3b_6\mathcal{H}\varphi''}{\varphi'^3}\partial_l\delta\varphi\partial_j\partial_k\delta\varphi
 \right.\right.
        \nonumber\\
       &\left.\left.
-\frac{3b_6\mathcal{H}}{\varphi'^2}\partial_l\delta\varphi'\partial_j\partial_k\delta\varphi
+\frac{12b_6 \mathcal{H}}{\varphi'}\partial_l\psi\partial_j\partial_k\delta\varphi
+\frac{3 b_6}{\varphi'}\partial_l\psi'\partial_j\partial_k\delta\varphi
 \right.\right.
        \nonumber\\
       &\left.\left.
-\frac{3b_{6X}\mathcal{H}}{\varphi'}\partial_l\delta X\partial_j\partial_k\delta\varphi
-\frac{3b_{6\varphi}\mathcal{H}}{\varphi'}\partial_l\delta\varphi\partial_j\partial_k\delta\varphi
+\frac{b_6}{\varphi'^2}\partial_l\nabla^2\delta\varphi\partial_j\partial_k\delta\varphi\right]
\right.\nonumber\\
&\left.
+\frac{3b_6\mathcal{H}}{\varphi'^2}\nabla^2(\partial_l\delta\varphi\partial_j\partial_k\delta\varphi)
\right\}h_{ij},
\end{align}
\begin{align}
S^{({\mathrm{PV7}})}_{ssh}=&\frac{1}{2\kappa^2} \int \mathrm{d}^4 x~\frac12\epsilon^{kl}_{\ \ i}\left\{-\partial_{\eta}^2\left(\frac{b_7}{\varphi'}\partial_l\delta\varphi\partial_j\partial_k\psi\right)
+\partial_{\eta}\Bigg[
\frac{b_7\mathcal{H}'}{\varphi'^2}\partial_l\delta\varphi\partial_j\partial_k\delta\varphi
 \right.
        \nonumber\\
       &\left.
-\frac{b_7\mathcal{H}^2}{\varphi'^2}\partial_l\delta\varphi\partial_j\partial_k\delta\varphi
+\frac{2b_7}{\varphi'}\partial_l\delta\varphi\partial_j\partial_k\psi'
+\frac{4b_7 \mathcal{H}}{\varphi'}
\partial_l\delta\varphi\partial_j\partial_k\psi
+b_7\partial_l\psi\partial_j\partial_k\psi
 \right.
        \nonumber\\
       &\left.
+b_{7\varphi}\partial_l\delta\varphi\partial_j\partial_k\psi
+b_{7X}\partial_l\delta X\partial_j\partial_k\psi\Bigg]+\frac{b_7}{\varphi'}\nabla^2(\partial_l\delta\varphi\partial_j\partial_k\psi)
\right\}h_{ij}.
\end{align}
where $b_{n\varphi}\equiv \partial b_n/\partial\varphi$, $b_{nX}\equiv \partial b_n/\partial X$ with $ X\equiv-\tfrac12\nabla_a\varphi\,\nabla^a\varphi$, and $\delta X=(\varphi'\delta\varphi'-\psi\varphi'^2)/a^2$. Here we have used $\phi=\psi$.

\section{EOM of the background and the linear scalar perturbations}\label{app-bf}

In this appendix, we present the equations of motion  for the background and the linear scalar perturbations.

For the background, the Friedmann equations are given by
\begin{align}
\label{EOM01}
3\mathcal{H}^2 &= \kappa^{2}a^2 \bar{\rho}, \\
\label{EOM02}
-\mathcal{H}^2 - 2\mathcal{H}^{\prime} &= \kappa^{2}a^2\bar{p},
\end{align}
where the background energy density and pressure are defined as
\begin{equation}
\bar{\rho} = \frac{1}{2a^2}{\varphi^{\prime}}^{2} + V(\varphi),
\qquad
\bar{p} = \frac{1}{2a^2}{\varphi^{\prime}}^{2} - V(\varphi).
\end{equation}

In the absence of anisotropic stress, the EOM for the linear scalar perturbations take the following form:
\begin{align}
\label{EOM100}
\nabla^2\psi - 3\mathcal{H}(\psi' + \mathcal{H}\phi)
&= \frac{1}{2}\kappa^{2}\left(\varphi^{\prime}{\delta \varphi}' - {\varphi^{\prime}}^2\psi + a^{2}V_{\varphi}\delta \varphi\right),\\
\label{EOM10i}
\psi' + \mathcal{H}\phi
&= \frac{1}{2}\kappa^{2}\varphi^{\prime}\delta \varphi, \\
\label{EOM1ij}
\psi'' + 2\mathcal{H}\psi' + \mathcal{H}\phi' + (\mathcal{H}^2 + 2\mathcal{H}')\phi
&= \frac{1}{2}\kappa^{2}\left(\varphi^{\prime}{\delta \varphi}' - {\varphi^{\prime}}^2\psi - a^{2}V_{\varphi}\delta \varphi\right),\\
\label{phi-psi}
\psi - \phi &= 0.
\end{align}

From the above equations, it is evident that the PV terms do not influence the evolution of either the background or the linear scalar perturbations.

\section{The  kernel }\label{app-kernel}
For the sake of clarity, we split $I^A$ defined in Eq. \eqref{IA}  into two parts as follows 
\begin{align}
    I^A(k,u,v,x)
=\frac{\sin x}{x}\left(I_{\mathrm{GRs}}+I^A_{\mathrm{PVs}}\right)
    +\frac{\cos x}{x}\left(I_{\mathrm{GRc}}+I^A_{\mathrm{PVc}}\right),
\end{align}
where the subscripts ''s" and ''c" stand for contributions involving the sine and cosine functions, respectively. 
We can write 
\begin{align}
I^A_{\mathrm{PVs}}&=
\mathcal{I}_{\mathrm{PVs}}(k, u,v,x)
- \mathcal{I}_{\mathrm{PVs}}(k,u,v,0),\nonumber\\
I^A_{\mathrm{PVc}}&=
\mathcal{I}_{\mathrm{PVc}}(k,u,v,x)
- \mathcal{I}_{\mathrm{PVc}}(k,u,v,0),
\end{align}
where ~$\mathcal{I}_{\mathrm{PVs}}$ and $\mathcal{I}_{\mathrm{PVc}}$ are defined by
\begin{gather}
\mathcal{I}_{\mathrm{PVs}}(u,v,y)=
\int \mathrm{d} y~ \cos (y) y\cdot f_{\mathrm{PV}}(k,u,v,y),\nonumber\\
\mathcal{I}_{\mathrm{PVc}}(u,v,y)=-
\int \mathrm{d} y ~\sin (y) y\cdot f_{\mathrm{PV}}(k,u,v,y).
\end{gather}

After tedious manipulations, the concrete expressions of  $\mathcal{I}_{\mathrm{PVs}}$ and $\mathcal{I}^A_{\mathrm{PVc}}$
are found to be
\begin{align}
\mathcal{I}_{\mathrm{PV3s}}=&\frac{\mathcal{C}_3\lambda^A k^4}{36u v y^3}\Bigg\{
24u v y^2 \cos y \cos \frac{u y}{\sqrt{3}} \cos \frac{v y}{\sqrt{3}}
-36y\sin y \sin \frac{u y}{\sqrt{3}} \sin \frac{v y}{\sqrt{3}}
\nonumber\\
&
-24 \sqrt{3} v y \cos y \sin \frac{u y}{\sqrt{3}}\cos \frac{v y}{\sqrt{3}}
-24 \sqrt{3} u y \cos y \cos \frac{u y}{\sqrt{3}}  \sin \frac{v y}{\sqrt{3}}\nonumber\\
&+12 \sqrt{3} v y^2  \sin y \sin \frac{u y}{\sqrt{3}} \cos \frac{v y}{\sqrt{3}}
+12 \sqrt{3} u y^2  \sin y \cos \frac{u y}{\sqrt{3}} \sin \frac{v y}{\sqrt{3}}\nonumber\\
&-12 \left[-6+ \left(3+u^2+ v^2\right)y^2\right] \cos y\sin \frac{u y}{\sqrt{3}} \sin \frac{v y}{\sqrt{3}}
\Bigg\}\nonumber\\
&+\frac{\mathcal{C}_3\lambda^Ak^4}{36u v }\Bigg\{9\left(-
\text{Si}\Big[\Big(1-\frac{v-u}{\sqrt{3}}\Big)y\Big]
- \text{Si}\Big[\Big(1+\frac{v-u}{\sqrt{3}}\Big)y\Big] \right.\nonumber\\
&\left.
+~\text{Si}\Big[\Big(1-\frac{v +u}{\sqrt{3}}\Big)y\Big]
+\text{Si}\Big[\Big(1+\frac{v +u}{\sqrt{3}}\Big)y\Big]
\right)\nonumber\\
&+\sqrt{3}u^3\left(-
\text{Si}\Big[\Big(1-\frac{v-u}{\sqrt{3}}\Big)y\Big]
+\text{Si}\Big[\Big(1+\frac{v-u}{\sqrt{3}}\Big)y\Big] \right.\nonumber\\
&\left.
-\text{Si}\Big[\Big(1-\frac{v +u}{\sqrt{3}}\Big)y\Big]
+\text{Si}\Big[\Big(1+\frac{v +u}{\sqrt{3}}\Big)y\Big]
\right)\nonumber\\
&+\sqrt{3}v^3\left(
\text{Si}\Big[\Big(1-\frac{v-u}{\sqrt{3}}\Big)y\Big]
-\text{Si}\Big[\Big(1+\frac{v-u}{\sqrt{3}}\Big)y\Big] \right.\nonumber\\
&\left. 
-\text{Si}\Big[\Big(1-\frac{v +u}{\sqrt{3}}\Big)y\Big]
+\text{Si}\Big[\Big(1+\frac{v +u}{\sqrt{3}}\Big)y\Big]
\right)\Bigg\},
\end{align}
\begin{align}
\mathcal{I}_{\mathrm{PV3c}}=&~\frac{\mathcal{C}_3\lambda^Ak^4}{36u v y^3}\Bigg\{-24 u v y^2 \sin y \cos \frac{u y}{\sqrt{3}}\cos \frac{v y}{\sqrt{3}}
-36 y  \cos y\sin \frac{u y}{\sqrt{3}}\sin \frac{v y}{\sqrt{3}}
\nonumber\\
&+24 \sqrt{3} v y \sin y \sin \frac{u y}{\sqrt{3}} \cos \frac{v y}{\sqrt{3}}
+24 \sqrt{3} u y \sin y \cos\frac{u y}{\sqrt{3}} \sin \frac{v y}{\sqrt{3}}
\nonumber\\
&+12\sqrt{3} v y^2  \cos y \sin \frac{u y}{\sqrt{3}} \cos \frac{v y}{\sqrt{3}}
+12\sqrt{3} u y^2 \cos y \cos \frac{u y}{\sqrt{3}} \sin \frac{v y}{\sqrt{3}}\nonumber\\
&+12  \left[-6+\left(3+ u^2+ v^2\right)y^2 \right] \sin y\sin  \frac{u y}{\sqrt{3}} \sin \frac{v y}{\sqrt{3}}
\Bigg\}\nonumber\\
&+\frac{\mathcal{C}_3\lambda^Ak^4}{36u v }\Bigg\{
9
\left(-\text{Ci}\Big[\Big(1-  \frac{v -u}{\sqrt{3}}\Big) y\Big]
-\text{Ci}\Big[\Big(1+\frac{v -u}{\sqrt{3}}\Big) y\Big] \right.\nonumber\\
&
\left.
+~\text{Ci}\Big[\Big|\Big(1- \frac{v +u}{\sqrt{3}}\Big) y\Big|\Big]
+\text{Ci}\Big[ \Big(1+ \frac{v+u}{\sqrt{3}} \Big)y\Big] 
\right) \nonumber\\
&+\sqrt{3}u^3\left(-\text{Ci}\Big[\Big(1-  \frac{v -u}{\sqrt{3}}\Big) y\Big]
+\text{Ci}\Big[\Big(1+\frac{v -u}{\sqrt{3}}\Big) y\Big] \right.\nonumber\\
&
\left.
-\text{Ci}\Big[\Big|\Big(1- \frac{v +u}{\sqrt{3}}\Big) y\Big|\Big]
+\text{Ci}\Big[ \Big(1+ \frac{v+u}{\sqrt{3}} \Big)y\Big] 
\right)\nonumber\\
&+\sqrt{3}v^3\left(\text{Ci}\Big[\Big(1-  \frac{v -u}{\sqrt{3}}\Big) y\Big]
-\text{Ci}\Big[\Big(1+\frac{v -u}{\sqrt{3}}\Big) y\Big] \right.\nonumber\\
&
\left.
-\text{Ci}\Big[\Big|\Big(1- \frac{v +u}{\sqrt{3}}\Big) y\Big|\Big]
+\text{Ci}\Big[ \Big(1+ \frac{v+u}{\sqrt{3}} \Big)y\Big] 
\right)\Bigg\},
\end{align}
and
\begin{align}
\mathcal{I}_{\mathrm{PV4s}}=&-\frac{\mathcal{C}_4\lambda^A k^4}{288u^3 v^3 y^3}\Bigg\{
24u v\left(u^2v^2+6(u^2+v^2)\right) y^2 \cos y \cos \frac{u y}{\sqrt{3}} \cos \frac{v y}{\sqrt{3}}\nonumber\\
&
-72y\left(2u^2v^2+3(u^2+v^2)\right)\sin y \sin \frac{u y}{\sqrt{3}} \sin \frac{v y}{\sqrt{3}}
\nonumber\\
&
-12 \sqrt{3} v y\left(5u^2v^2+12(u^2+v^2)\right) \cos y \sin \frac{u y}{\sqrt{3}}\cos \frac{v y}{\sqrt{3}}
\nonumber\\
&
-12 \sqrt{3} u y\left(5u^2v^2+12(u^2+v^2)\right) \cos y \cos \frac{u y}{\sqrt{3}}  \sin \frac{v y}{\sqrt{3}}\nonumber\\
&+12 \sqrt{3} v y^2\left(u^2v^2+6(u^2+v^2)\right)  \sin y \sin \frac{u y}{\sqrt{3}} \cos \frac{v y}{\sqrt{3}}\nonumber\\
&
+12 \sqrt{3} u y^2 \left(u^2v^2+6(u^2+v^2)\right) \sin y \cos \frac{u y}{\sqrt{3}} \sin \frac{v y}{\sqrt{3}}\nonumber\\
&-12 \Big[-12(3u^2+3v^2+2u^2v^2)+u^2v^2(u^2+v^2)y^2\nonumber\\
&+6(3u^2+3v^2-2u^4-2v^4+4u^2v^2)y^2\Big] \cos y\sin \frac{u y}{\sqrt{3}} \sin \frac{v y}{\sqrt{3}}
\Bigg\}\nonumber\\
&-\frac{\mathcal{C}_4\lambda^Ak^4}{288u^3 v^3 }\Bigg\{18\left(3(u^2+v^2)-3 (u^4+v^4)+2 u^2 v^2\right)\left(-
\text{Si}\Big[\Big(1-\frac{v-u}{\sqrt{3}}\Big)y\Big]
 \right.\nonumber\\
&\left.
- \text{Si}\Big[\Big(1+\frac{v-u}{\sqrt{3}}\Big)y\Big]
+~\text{Si}\Big[\Big(1-\frac{v +u}{\sqrt{3}}\Big)y\Big]
+\text{Si}\Big[\Big(1+\frac{v +u}{\sqrt{3}}\Big)y\Big]
\right)\nonumber\\
&+\sqrt{3}u^3\left( u^2 \left(v^2-12\right)+3 v^2 \left(v^2+5\right)\right)\left(-
\text{Si}\Big[\Big(1-\frac{v-u}{\sqrt{3}}\Big)y\Big]
 \right.\nonumber\\
&\left.
+\text{Si}\Big[\Big(1+\frac{v-u}{\sqrt{3}}\Big)y\Big]
-\text{Si}\Big[\Big(1-\frac{v +u}{\sqrt{3}}\Big)y\Big]
+\text{Si}\Big[\Big(1+\frac{v +u}{\sqrt{3}}\Big)y\Big]
\right)\nonumber\\
&+\sqrt{3}v^3 \left(v^2(u^2 -12)+3u^2(u^2+5) \right)\left(
\text{Si}\Big[\Big(1-\frac{v-u}{\sqrt{3}}\Big)y\Big]
 \right.\nonumber\\
&\left. 
-\text{Si}\Big[\Big(1+\frac{v-u}{\sqrt{3}}\Big)y\Big]
-\text{Si}\Big[\Big(1-\frac{v +u}{\sqrt{3}}\Big)y\Big]
+\text{Si}\Big[\Big(1+\frac{v +u}{\sqrt{3}}\Big)y\Big]
\right)\Bigg\},
\end{align}
\begin{align}
\mathcal{I}_{\mathrm{PV4c}}=&~\frac{\mathcal{C}_4\lambda^A k^4}{288u^3 v^3 y^3}\Bigg\{
24u v\left(u^2v^2+6(u^2+v^2)\right) y^2\sin y \cos \frac{u y}{\sqrt{3}}\cos \frac{v y}{\sqrt{3}}\nonumber\\
&
+72y\left(2u^2v^2+3(u^2+v^2)\right)\cos y\sin \frac{u y}{\sqrt{3}}\sin \frac{v y}{\sqrt{3}}
\nonumber\\
&-12 \sqrt{3} v y\left(5u^2v^2+12(u^2+v^2)\right)  \sin y \sin \frac{u y}{\sqrt{3}} \cos \frac{v y}{\sqrt{3}}\nonumber\\
&
-12 \sqrt{3} u y\left(5u^2v^2+12(u^2+v^2)\right) \sin y \cos\frac{u y}{\sqrt{3}} \sin \frac{v y}{\sqrt{3}}
\nonumber\\
&-12 \sqrt{3} v y^2\left(u^2v^2+6(u^2+v^2)\right)  \cos y \sin \frac{u y}{\sqrt{3}} \cos \frac{v y}{\sqrt{3}}\nonumber\\
&
-12 \sqrt{3} u y^2 \left(u^2v^2+6(u^2+v^2)\right) \cos y \cos \frac{u y}{\sqrt{3}} \sin \frac{v y}{\sqrt{3}}\nonumber\\
&-12 \Big[-12(3u^2+3v^2+2u^2v^2)+u^2v^2(u^2+v^2)y^2\nonumber\\
&+6(3u^2+3v^2-2u^4-2v^4+4u^2v^2)y^2\Big] \sin y\sin  \frac{u y}{\sqrt{3}} \sin \frac{v y}{\sqrt{3}}
\Bigg\}\nonumber\\
&-\frac{\mathcal{C}_4\lambda^Ak^4}{288u^3 v^3 }\Bigg\{18\left(3(u^2+v^2)-3 (u^4+v^4)+2 u^2 v^2\right)\left(-
\text{Ci}\Big[\Big(1-\frac{v-u}{\sqrt{3}}\Big)y\Big]
 \right.\nonumber\\
&\left.
- \text{Ci}\Big[\Big(1+\frac{v-u}{\sqrt{3}}\Big)y\Big]
+~\text{Ci}\Big[\Big(1-\frac{v +u}{\sqrt{3}}\Big)y\Big]
+\text{Ci}\Big[\Big(1+\frac{v +u}{\sqrt{3}}\Big)y\Big]
\right)\nonumber\\
&+\sqrt{3}u^3\left( u^2 \left(v^2-12\right)+3 v^2 \left(v^2+5\right)\right)\left(-
\text{Ci}\Big[\Big(1-\frac{v-u}{\sqrt{3}}\Big)y\Big]
 \right.\nonumber\\
&\left.
+\text{Ci}\Big[\Big(1+\frac{v-u}{\sqrt{3}}\Big)y\Big]
-\text{Ci}\Big[\Big(1-\frac{v +u}{\sqrt{3}}\Big)y\Big]
+\text{Ci}\Big[\Big(1+\frac{v +u}{\sqrt{3}}\Big)y\Big]
\right)\nonumber\\
&+\sqrt{3}v^3 \left(v^2(u^2 -12)+3u^2(u^2+5) \right)\left(
\text{Ci}\Big[\Big(1-\frac{v-u}{\sqrt{3}}\Big)y\Big]
 \right.\nonumber\\
&\left. 
-\text{Ci}\Big[\Big(1+\frac{v-u}{\sqrt{3}}\Big)y\Big]
-\text{Ci}\Big[\Big(1-\frac{v +u}{\sqrt{3}}\Big)y\Big]
+\text{Ci}\Big[\Big(1+\frac{v +u}{\sqrt{3}}\Big)y\Big]
\right)\Bigg\},
\end{align}

where
\begin{equation}
\operatorname{Si}(x)=\int_0^x \mathrm{~d} y \frac{\sin y}{y}, \quad \operatorname{Ci}(x)=-\int_x^{\infty} \mathrm{d} y \frac{\cos y}{y}  .
\end{equation}

Since we are interested in the SIGWs at the present time, we take  $x\gg 1$. In this limit, we have
\begin{align}
    I_{\mathrm{PV3s}}^A(k, u, v, x\rightarrow\infty)=-\frac{\mathcal C_3\lambda^Ak^4}{36u v}\left(9-\sqrt{3}(u^3+v^3)\right)\pi~\Theta(v+u-\sqrt{3}),~~~~~~~~~~~
\end{align}
\begin{align}
            I_{\mathrm{PV3c}}^A(k, u, v, x\rightarrow\infty)=&-\frac{\mathcal C_3\lambda^Ak^4}{36u v}\Bigg(12uv+9\log \left|\frac{3-(u+v)^2}{3-(u-v)^2}\right|+\sqrt{3}u^3\log \left|\frac{(\sqrt{3}+v)^2-u^2}{(\sqrt{3}-v)^2-u^2}\right|\nonumber\\
& +\sqrt{3}v^3\log \left|\frac{(\sqrt{3}+u)^2-v^2}{(\sqrt{3}-u)^2-v^2}\right| \Bigg).
\end{align}
Similarly, for the other terms, we have
\begin{equation}
    \begin{split}
I_{\mathrm{PV4s}}^A(k, u, v, x\rightarrow\infty)=&\frac{\mathcal C_4\lambda^Ak^4}{288u^3 v^3}\Bigg[18\Big(3(u^2+v^2)+2u^2v^2-3(u^4+v^4)\Big)\\
&-\sqrt{3}u^3 \Big(u^2 \left(v^2-12\right)+3  v^2 \left(v^2+5\right)\Big)\\
& -\sqrt{3}v^3 \Big(v^2(u^2-12)+3u^2(u^2+5)\Big)\Bigg]\pi~\Theta(v+u-\sqrt{3}),
    \end{split}
\end{equation}
\begin{equation}
    \begin{split}
        I_{\mathrm{PV4c}}^A(k, u, v, x\rightarrow\infty)=&-\frac{\mathcal C_4\lambda^Ak^4}{288u^3 v^3}\Bigg[24 u^3 v^3-72 u^3 v-72 u v^3\\
        &-18\Big(3(u^2+v^2)-3(u^4+v^4)+2u^2v^2\Big)\log \Big|\frac{3-(u+v)^2}{3-(u-v)^2}\Big|\\
        &-\sqrt{3}u^3 \Big(u^2 \left(v^2-12\right)+3  v^2 \left(v^2+5\right)\Big)\log \Big|\frac{(\sqrt{3}+v)^2-u^2}{(\sqrt{3}-v)^2-u^2}\Big|\\
& -\sqrt{3}v^3 \Big(v^2(u^2-12)+3u^2(u^2+5)\Big)\log \Big|\frac{(\sqrt{3}+u)^2-v^2}{(\sqrt{3}-u)^2-v^2}\Big| \Bigg],
    \end{split}
\end{equation}

and
\begin{equation}
    \begin{split}
I_{\mathrm{GRs}}(u, v, x\rightarrow\infty)=&\frac{3\left(u^2+v^2-3\right)}{4 u^3 v^3}\left(-4 u v+\left(u^2+v^2-3\right) \log \left|\frac{3-(u+v)^2}{3-(u-v)^2}\right|\right),\\
I_{\mathrm{GRc}}( u, v, x\rightarrow\infty)=&\frac{3\left(u^2+v^2-3\right)}{4 u^3 v^3}\left(-\pi\left(u^2+v^2-3\right) \Theta(v+u-\sqrt{3})\right).
    \end{split}
\end{equation}
As a result, the time average is
\begin{equation}
    \begin{split}
        &\overline{I^A(k, u, v, x \rightarrow \infty)^2}\\
=&
\frac{1}{2 x^2}\Big\{\Big(I_{\mathrm{GRs}}(u, v, x\rightarrow \infty)+I_{\mathrm{PV3s}}(u, v, x\rightarrow \infty)+I_{\mathrm{PV4s}}(u, v, x\rightarrow \infty)\Big)^2\\
&+\Big(I_{\mathrm{GRc}}(u, v, x\rightarrow \infty)+I_{\mathrm{PV3c}}(u, v, x\rightarrow \infty)+I_{\mathrm{PV4c}}(u, v, x\rightarrow \infty)\Big)^2
\Big\}.
    \end{split}
\end{equation}

\newpage
\section{log-normal case}\label{fig-log-normal}
\begin{figure}[!htbp]
  \centering
\includegraphics[width=0.95\textwidth]{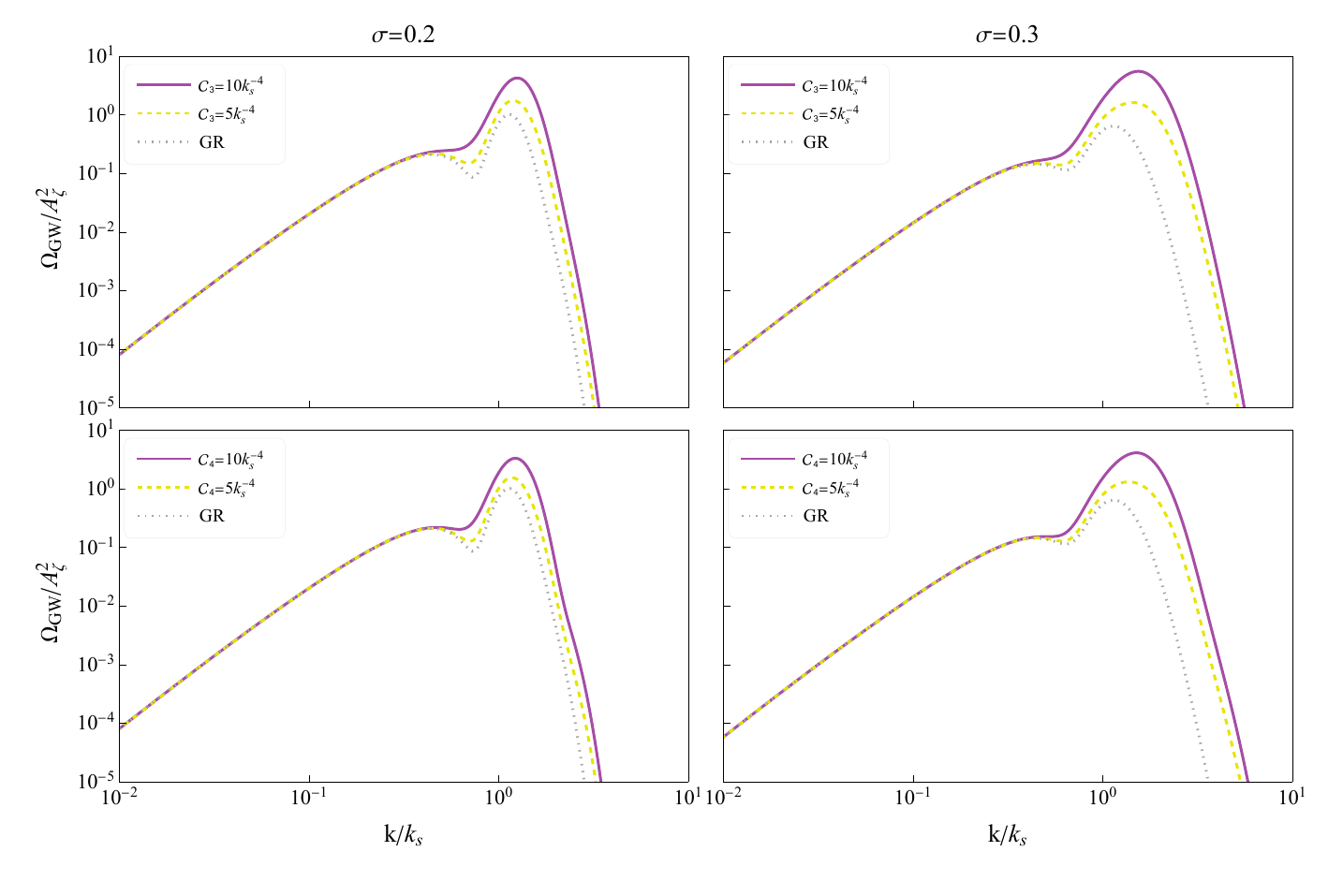}
\caption{The fractional energy density $\Omega_{\mathrm{GW}}$ of SIGWs in GR and the PVST theory with a log-normal spectrum. Left: $\sigma=0.2$.
Right: $\sigma=0.3$.}\label{LfinalPlotOM}
\end{figure}

\begin{figure}[!htbp]
  \centering
\includegraphics[width=0.95\textwidth]{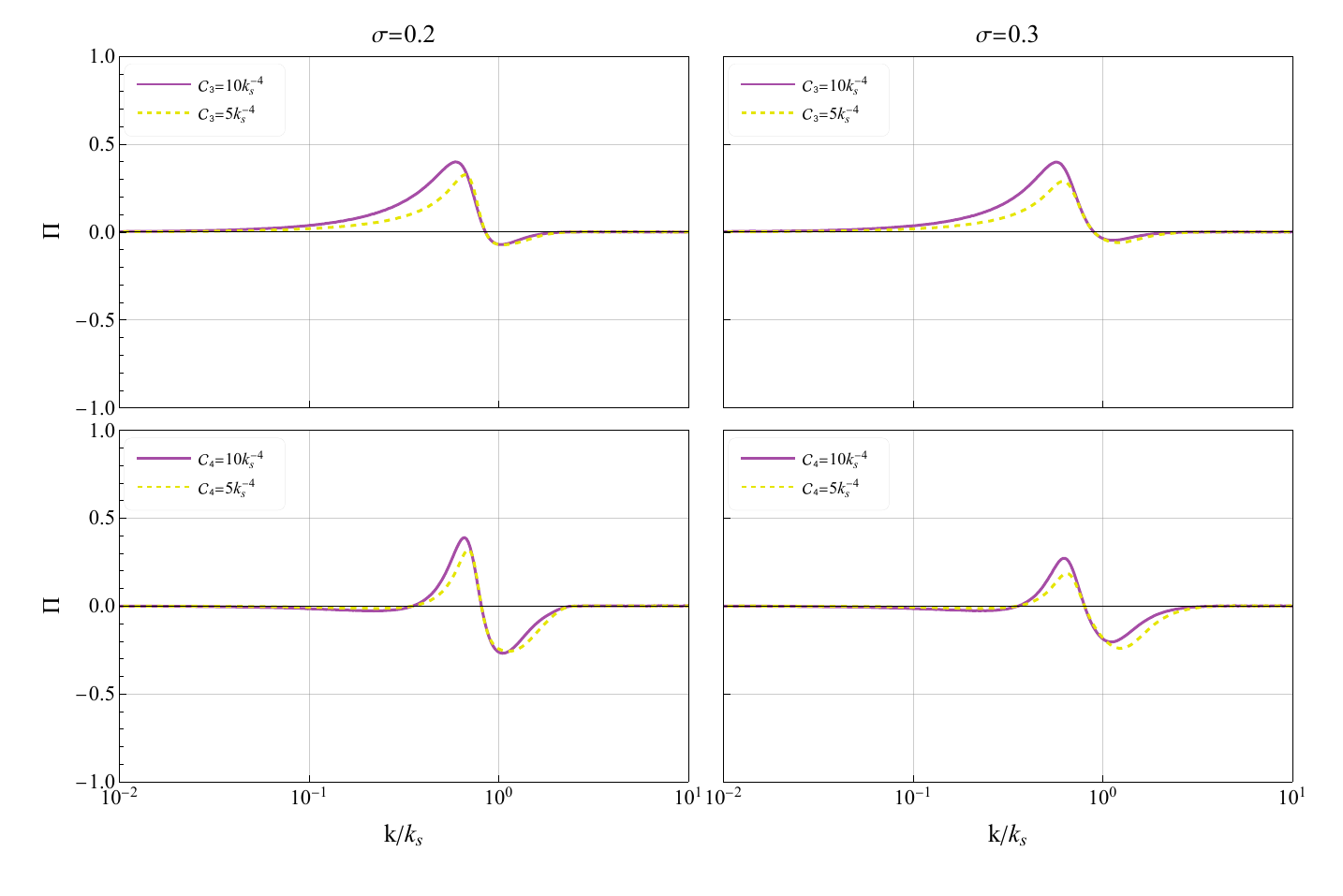}
\caption{The degree of circular polarization of SIGWs with a log-normal spectrum. Left: $\sigma=0.2$.
 Right: $\sigma=0.3$.}\label{LfinalPlotPI}
\end{figure}



\clearpage

\end{document}